\documentclass[12pt,a4paper,aps,preprint,superscriptaddress,nofootinbib]{revtex4-1}
\usepackage[utf8]{inputenc}
\usepackage{graphicx}
\usepackage{amssymb}
\usepackage{textcomp}
\usepackage{amsmath}
\usepackage{tabularx}
\usepackage{bm}
\usepackage{times}
\usepackage{color}
\usepackage{slashed}
\usepackage{multirow}
\usepackage{verbatim}
\usepackage{cancel}
\usepackage{subfigure}
\usepackage{ulem}

\usepackage[colorlinks=true, pdfstartview=FitV, linkcolor=blue, citecolor=blue, urlcolor=blue]{hyperref}
\allowdisplaybreaks[4]

\def\lsim{\mathrel{\raise.3ex\hbox{$<$\kern-.75em\lower1ex\hbox{$\sim$}}}}
\def\gsim{\mathrel{\raise.3ex\hbox{$>$\kern-.75em\lower1ex\hbox{$\sim$}}}}

\definecolor{orange}{rgb}{1,0.5,0}

\begin{document}

\title{Quantum calculation of axion-photon transition in electromagnetodynamics for cavity haloscope}

\author{Tong Li}
\email{litong@nankai.edu.cn}
\author{Rui-Jia Zhang}
\email{zhangruijia@mail.nankai.edu.cn}
\affiliation{
School of Physics, Nankai University, Tianjin 300071, China
}

\begin{abstract}
The Witten effect implies the presence of electric charge of magnetic monople and the possible relationship between axion and dyon. The axion-dyon dynamics can be reliably built based on the quantum electromagnetodynamics (QEMD) which was developed by Schwinger and Zwanziger in 1960's.
A generic low-energy axion-photon effective field theory can also be realized in the language of ``generalized symmetries'' with higher-form symmetries and background gauge fields.
In this work, we implement the quantum calculation of the axion-single photon transition rate inside a homogeneous electromagnetic field in terms of the new axion interaction Hamiltonian in QEMD. This quantum calculation can clearly imply the enhancement of conversion rate through resonant cavity in axion haloscope experiments. We also show the promising potentials on the cavity search of new axion-photon couplings.
\end{abstract}

\maketitle

\section{Introduction}
\label{sec:Intro}

It is well-known that the strong CP problem in quantum chromodynamics (QCD) arises from the source of CP violation in QCD Lagrangian with $\theta G^{a~\mu\nu}\tilde{G}^a_{\mu\nu}$. The Peccei-Quinn (PQ) mechanism solves the strong CP problem by introducing a pseudo-Goldstone boson $a$ called axion after the spontaneous breaking of a QCD anomalous $U(1)_{\rm PQ}$ global symmetry~\cite{Peccei:1977hh,Peccei:1977ur,Weinberg:1977ma,Wilczek:1977pj,Baluni:1978rf,Crewther:1979pi,Kim:1979if,Shifman:1979if,Dine:1981rt,Zhitnitsky:1980tq,Baker:2006ts,Pendlebury:2015lrz}. The chiral transformation of the quark fields with PQ charges also leads to the anomaly under QED and the coupling $g_{a\gamma\gamma}aF^{\mu\nu}\tilde{F}_{\mu\nu}$ between axion and electromagnetic fields. In 1979, E.~Witten showed that a CP violating term $\theta F^{\mu\nu}\tilde{F}_{\mu\nu}$ with non-zero vacuum angle $\theta$ provides an electric charge $-\theta e/2\pi$ for magnetic monopoles~\cite{Witten:1979ey}. This so-called
Witten effect implies a close relationship between axion and magnetic monopole due to the axion-photon coupling $g_{a\gamma\gamma}a \vec{\mathbb{E}}\cdot \vec{\mathbb{B}}$.

This axion-dyon dynamics was first derived by W.~Fischler et al. under the classical electromagnetism~\cite{Fischler:1983sc} and was proposed to solve cosmological problems in recent years~\cite{Kawasaki:2015lpf,Nomura:2015xil,Kawasaki:2017xwt,Houston:2017kwe,Sato:2018nqy}.
In their works, however, the magnetic monopoles were treated as quasi-classical external sources and the quantization of electromagnetism is not complete. A reliable quantization in the presence of magnetic monopoles was developed by J.~S.~Schwinger and D.~Zwanziger in 1960's and called quantum electromagnetodynamics (QEMD)~\cite{Schwinger:1966nj,Zwanziger:1968rs,Zwanziger:1970hk}. Recently, based on the QEMD framework, Ref.~\cite{Sokolov:2022fvs} constructed a more generic axion-photon Lagrangian in the low-energy axion effective field theory (EFT).
Besides the Witten effect term, more anomalous axion-photon interactions and couplings arise assuming the existence of heavy PQ-charged fermions with electric
and magnetic charges. This is in contrary to the conventional axion EFT $g_{a\gamma\gamma} aF^{\mu\nu}\tilde{F}_{\mu\nu}$ in the quantum electrodynamics (QED) framework. As a result of the above generic axion-photon Lagrangian, the conventional axion Maxwell equations~\cite{Sikivie:1983ip} are further modified and some consequent new detection strategies of axion are studied in recent years~\cite{Li:2022oel,Tobar:2022rko,McAllister:2022ibe,Li:2023kfh}.

The key property of QEMD is to substitute the $U(1)_{\rm EM}$ gauge group in the Standard Model (SM) by two $U(1)$ gauge groups $U(1)_{\rm E}\times U(1)_{\rm M}$ to introduce both electric and magnetic charges. Then, two four-potentials $A^\mu$ and $B^\mu$ (instead of only one in QED) are introduced corresponding to the two $U(1)$ gauge groups, respectively. A non-trivial form of equal-time canonical commutation relations between them can be built~\cite{Zwanziger:1970hk}. It guarantees the preservation of the right degrees of freedom of physical photon.

The other property of QEMD is that it seemingly acts like a non-local quantum field theory (QFT). To obtain the covariant Maxwell equations in the presence of conserved electric current $j_e$ and magnetic current $j_m$, one needs to introduce an arbitrary spacelike vector $n^\mu=(0,\vec{n})$. The electromagnetic field strength tensor $F_{\mu\nu}$ and its dual tensor $F_{\mu\nu }^d$~\footnote{Below we define $X^d_{\mu\nu}\equiv\tilde{X}_{\mu\nu}=\epsilon_{\mu\nu\alpha\beta}X^{\alpha\beta}/2$ as the Hodge dual of tensor $X_{\mu\nu}$. Also, $(Y \wedge Z)^{\mu \nu}\equiv Y^\mu Z^\nu - Y^\nu Z^\mu$ for any four-vectors $Y$ and $Z$.} are given in this way
\begin{eqnarray}
F=\partial \wedge A - (n\cdot \partial)^{-1} (n\wedge j_m)^d\;,~~
F^d=\partial \wedge B + (n\cdot \partial)^{-1} (n\wedge j_e)^d\;,
\label{eq:FFd}
\end{eqnarray}
where the integral operator $(n\cdot \partial)^{-1}$ satisfies $n\cdot \partial (n\cdot \partial)^{-1}(\vec{x})=\delta(\vec{x})$.
The second terms on the right-handed sides of Eq.~(\ref{eq:FFd}) induce likely non-local property in QEMD.
One can prove that the non-local part does not play any role in physical processes and the Lorentz invariance is not violated~\cite{Brandt:1977be,Brandt:1978wc,Sokolov:2023pos}.

The QCD axion (see Ref.~\cite{DiLuzio:2020wdo} for a recent review) can become as dark matter (DM) candidate through the misalignment mechanism~\cite{Preskill:1982cy,Dine:1982ah}.
The conventional axion haloscope experiments such as ADMX~\cite{ADMX:2018gho,ADMX:2021nhd} are built based on resonant cavity technique to search for $\mathcal{O}(10)~\mu{\rm eV}$ axion DM. The cosmic axions resonantly convert into a monochromatic photon with the enhancement from a high quality factor $Q$ when the cavity resonant frequency is tuned to the axion mass $m_a$. The mean number of thermal photons in the cavity at finite temperature $T$ is given by
\begin{eqnarray}
n(\omega_a,T)={1\over e^{\omega_a/k_B T}-1}\;,
\end{eqnarray}
where $k_B$ is Boltzmann constant.
When $m_a\gtrsim \mathcal{O}(10)~\mu{\rm eV}$ and $T\approx 20~{\rm mK}$, the occupation number of thermal photon is quite low and the cavity can be treated as a single photon emitter. Although the usual electromagnetic power radiated in cavity is calculated in classical field theory~\cite{Sikivie:1983ip}, the actual description is quantum mechanical process of axion to photon conversion as stated by P.~Sikivie in Ref.~\cite{Sikivie:2020zpn}.
To describe this axion-single photon conversion $a\to \gamma$, the calculation of transition rate should be performed at quantum level~\cite{Beutter:2018xfx,Yang:2022uil}.
In this work, we follow Ref.~\cite{Yang:2022uil} to implement the quantum calculation of photon $|0\rangle \to |1\rangle $ transition rate inside a homogeneous electromagnetic field in terms of the new axion interaction Hamiltonian based on QEMD. This quantum calculation can clearly imply the enhancement of conversion rate through resonance which is not certain in the classical picture. Our work will show the basic method for the generic cavity search of the new axion-photon couplings.

This paper is organized as follows. In Sec.~\ref{sec:genericaxion}, we introduce the generic axion-photon interactions. We will show the realizations of this theory in both QEMD and generalized symmetry. In Sec.~\ref{sec:Photon}, we perform a complete quantum calculation of $a\to \gamma$ transition rate based on the new axion interaction Hamiltonian in QEMD. The transition rates from different types of cavity mode are obtained under external static magnetic or electric background. We also show the sensitivity of resonant cavity to axion-photon couplings in Sec.~\ref{sec:Cavity}. Our conclusions are drawn in Sec.~\ref{sec:Con}.

\section{The realizations of generic axion-photon interactions in QEMD and generalized symmetry}
\label{sec:genericaxion}

\subsection{The generic axion-photon interactions in QEMD}
\label{sec:QEMD}

In the QEMD theory, the photon is described by two four-potentials $A^\mu$ and $B^\mu$. Correspondingly, the gauge group of QEMD becomes $U(1)_{\rm E}\times U(1)_{\rm M}$ which inherently introduces both electric and magnetic charges.
The equal-time canonical commutation relations between the two four-potentials were obtained~\cite{Zwanziger:1970hk}
\begin{eqnarray}
\label{eq:commutation1}[A^\mu(t,\vec{x}),B^\nu(t,\vec{y})]&=&i\epsilon^{\mu\nu}_{~~\kappa 0} n^\kappa (n\cdot \partial)^{-1}(\vec{x}-\vec{y})\;,\\
~
\label{eq:commutation2}[A^\mu(t,\vec{x}),A^\nu(t,\vec{y})]&=&[B^\mu(t,\vec{x}),B^\nu(t,\vec{y})]=-i(g_0^{~\mu} n^\nu+g_0^{~\nu} n^\mu)(n\cdot \partial)^{-1}(\vec{x}-\vec{y})\;.
\end{eqnarray}
The electromagnetic field strength tensors $F$ and $F^{d}$ are then introduced in the way that
\begin{eqnarray}
n\cdot F=n\cdot (\partial \wedge A)\;,~~ n\cdot F^d=n\cdot (\partial \wedge B)\;,
\end{eqnarray}
where $n^\mu = (0,\vec{n})$ is an arbitrary fixed spatial vector. Apparently, the two four-potentials have opposite parities. In the absence of electric and magnetic currents $j_e, j_m$, one has a simplified form
\begin{eqnarray}
F = \partial \wedge A = -(\partial \wedge B)^d\;,~~~F^d =  \partial \wedge B = (\partial \wedge A)^d\;.
\end{eqnarray}
The $n$-related terms produce the non-locality in this theory. This non-local property can be realized by the two-particle irreducible representation in the QFT theory with both electric charge $q$ and magnetic charge $g$. Each two-particle state $(i,j)$ is characterized by the Dirac-Schwinger-Zwanziger (DSZ) quantization condition
\begin{eqnarray}
q_i g_j - q_j g_i=2\pi N\;,~~~N\in \mathbb{Z}\;.
\end{eqnarray}
Thus, the cluster decomposition principle is obviously violated by irreducible two-particle state~\cite{Sokolov:2023pos} and the Lorentz invariance is seemingly violated in this QEMD theory. However, it was formally shown that the observables of the QEMD are Lorentz invariant using the path-integral approach~\cite{Brandt:1977be,Brandt:1978wc,Sokolov:2023pos}. After all the quantum
corrections are properly accounted for, the dependence on the spatial vector $n_\mu$ in the action
$S$ factorizes into an integer linking number $L_n$ multiplied by the
combination of charges in the DSZ quantization condition $q_i g_j - q_j g_i$. This $n$ dependent part is then given by $2\pi N$ with $N$ being an integer. Since $S$ contributes to the generating functional in the exponential form $e^{iS}$, this
Lorentz-violating part does not play any role in physical processes.

The Lagrangian for the anomalous interactions between axion $a$ and photon in QEMD is given by~\cite{Sokolov:2022fvs}
\begin{eqnarray}
\mathcal{L}&\supset&  -{1\over 4} g_{aAA}~a~{\rm tr}[(\partial \wedge A)(\partial \wedge A)^d] - {1\over 4} g_{aBB}~a~{\rm tr}[(\partial \wedge B)(\partial \wedge B)^d]\nonumber \\
&& - {1\over 2} g_{aAB}~a~{\rm tr}[(\partial \wedge A)(\partial \wedge B)^d]  \;.
\label{eq:axion}
\end{eqnarray}
The first two dimension-five operators are CP-conserving axion interactions. Their couplings $g_{aAA}$ and $g_{aBB}$ are governed by the $U(1)_{\rm PQ}U(1)_{\rm E}^2$ and $U(1)_{\rm PQ}U(1)_{\rm M}^2$ anomalies, respectively. As $A^\mu$ and $B^\mu$ have opposite parities, the third operator is CP-violating one and its coupling $g_{aAB}$ is determined by the $U(1)_{\rm PQ}U(1)_{\rm E}U(1)_{\rm M}$ anomaly. The inclusion of this term accounts for the intrinsic CP violation in a dyon theory. It is analogous to the interaction between electromagnetic field and a scalar $\phi$ with positive parity $\phi F^{\mu\nu}F_{\mu\nu}$~\cite{Donohue:2021jbv}.
In terms of classical electromagnetic fields, the above axion-photon Lagrangian becomes~\footnote{We use symbol ``$\mathbb{B}$'' rather than ``$B$'' to denote magnetic field in order not to conflict with the four-potential $B^\mu$. To be consistent with it, we also use ``$\mathbb{E}$'' to denote electric field.}
\begin{eqnarray}
\mathcal{L}&\supset& -{1\over 4}(g_{aAA}-g_{aBB})~a~ F_{\mu\nu} F^{d~\mu\nu} + {1\over 2} g_{aAB}~a~ F_{\mu\nu} F^{\mu\nu}\nonumber \\
&=&(g_{aAA}-g_{aBB})~a~ \vec{\mathbb{B}}\cdot \vec{\mathbb{E}} + g_{aAB}~a~ (\vec{\mathbb{B}}^2-\vec{\mathbb{E}}^2)\;.
\label{eq:Lagrangian}
\end{eqnarray}
Note that the QEMD theory has an intrinsic source of CP violation. It is because the spectrum of dyon charges is not CP invariant with only a state $(q, g)$ and without its CP conjugate state $(-q, g)$. The intrinsic CP violation of high energy
QEMD is transferred to the low-energy axion-photon EFT after integrating out heavy fermionic dyons with charges $(q,g)$. The coefficient $g_{aAB}$ is determined by the CP violating anomaly coefficient and the $g_{aAB}$ term in the Lagrangian is a CP-odd term. They reflect the intrinsic CP violation of QEMD.
This is the form of interactions that we will use for the quantum calculation of $a\to \gamma$ transition rate below.
Taking care of the above anomalies, one can calculate the coupling coefficients as
\begin{eqnarray}
g_{aAA}={Ee^2\over 4\pi^2 v_{\rm PQ}}\;,~~g_{aBB}={Mg_0^2\over 4\pi^2 v_{\rm PQ}}\;,~~g_{aAB}={Deg_0\over 4\pi^2 v_{\rm PQ}}\;,
\label{eq:couplings}
\end{eqnarray}
where $e$ is the unit of electric charge, $g_0$ is the minimal magnetic charge with $g_0=2\pi/e$ in the DSZ quantization condition, and $v_{\rm PQ}$ is the $U(1)_{\rm PQ}$ symmetry breaking scale. $E(M)$ is the electric (magnetic) anomaly coefficient and $D$ is the mixed electric-magnetic CP-violating anomaly coefficient. They can be computed by integrating out heavy PQ-charged fermions with electric and magnetic charges. Ref.~\cite{Sokolov:2022fvs} performed the calculation of the anomaly coefficients by following Fujikawa's path integral method~\cite{Fujikawa:1979ay}. As the DSZ quantization condition indicates $g_0\gg e$, we have the scaling of the axion-photon couplings as $g_{aBB}\gg |g_{aAB}|\gg g_{aAA}$.

We can propose a KSVZ-like high energy QEMD theory with heavy fermions $\psi$ as a UV completion~\cite{Sokolov:2022fvs}. The
Lagrangian for the fermions $\psi$ is
\begin{eqnarray}
\mathcal{L}\supset i\bar{\psi} \gamma^\mu D_\mu \psi + y \Phi \bar{\psi}_L \psi_R +h.c.\;,
\end{eqnarray}
where $D_\mu=\partial_\mu -e q_\psi A_\mu -g_0 g_\psi B_\mu$ denotes a covariant
derivative with both $A_\mu$ and $B_\mu$ four-potentials multiplied by the corresponding
electric and magnetic charges, and $\Phi$ is the PQ complex scalar field. The coefficient $g_{aAB}$ can be obtained by integrating out the heavy dyons.

\subsection{Generalized symmetry realization}
\label{sec:generalsymmetry}

The QEMD theory describes monopole dynamics and more axion couplings arise based on QEMD as seen in last section. On the other hand, new axion couplings can also be realized in the language of higher-form symmetries in a topological QFT (TQFT)~\cite{Kapustin:2014gua,Seiberg:2010qd,Gaiotto:2014kfa}. The generic axion couplings then naturally arise when an axion-Maxwell theory couples to a TQFT~\cite{Brennan:2023kpw}. Below we briefly review the spirit of generalized symmetry~\cite{Kapustin:2014gua,Seiberg:2010qd,Gaiotto:2014kfa} and the realization of generic axion couplings in TQFT.

We consider a general $p$-form symmetry in $d$ dimensions. The symmetry transformation as an operator is associated with a co-dimension $p+1$ manifold $M^{(d-p-1)}$
\begin{eqnarray}
U_g(M^{(d-p-1)})\;,
\end{eqnarray}
where $g$ is an element of symmetry group $G$.
The operators obey the multiplication rule
\begin{eqnarray}
U_g(M^{(d-p-1)})U_{g'}(M^{(d-p-1)})=U_{g^{\prime\prime}}(M^{(d-p-1)})\;,
\end{eqnarray}
where $g^{\prime\prime}=gg^\prime\in G$.
The dependence of operator $U_g(M^{(d-p-1)})$ on manifold $M^{(d-p-1)}$ is topological and remains unchanged unless the deformation of $M^{(d-p-1)}$ crosses an operator $V$. The topological operator $U_g(M^{(d-p-1)})$ acts on a $p$-dimensional operator $V$ of manifold $C^{(p)}$ in the form of~\cite{Gaiotto:2014kfa}
\begin{eqnarray}
U_g(M^{(d-p-1)}) V(C^{(p)})= g(V)^{\langle M^{(d-p-1)},C^{(p)} \rangle} V(C^{(p)}) U_g(M^{(d-p-1)})\;,
\end{eqnarray}
where $g(V)$ is the representation of group element $g$ of $V$, and $\langle M^{(d-p-1)},C^{(p)} \rangle$ is the linking number for manifolds $M^{(d-p-1)}$ and $C^{(p)}$. It is then natural to couple the system to flat background gauge field of the higher-form symmetry.
Taking a $d=4$ Maxwell theory for illustration, one claims there are two one-form $U(1)$ symmetries, i.e. $U(1)_{\rm E}\times U(1)_{\rm M}$. The symmetries are generated by the integral of $d-1=3$-form currents
\begin{eqnarray}
U^{\rm E}(M^{(2)})&=&e^{i\alpha\oint_{M^{(2)}}j_e}\;,~~g_{\rm E}=e^{i\alpha}\in U(1)_{\rm E}\;,\\
U^{\rm M}(M^{(2)})&=&e^{i\beta\oint_{M^{(2)}}j_m}\;,~~g_{\rm M}=e^{i\beta}\in U(1)_{\rm M}\;,
\end{eqnarray}
where $g_{\rm E(M)}$ is the element of group $U(1)_{\rm E(M)}$, and the electric and magnetic charges are given by $q(M^{(2)})=\oint_{M^{(2)}}j_e$ and $g(M^{(2)})=\oint_{M^{(2)}}j_m$, respectively.
The representation then generally becomes
\begin{eqnarray}
(g_{\rm E(M)})^{Q \langle M^{(2)},C^{(1)} \rangle}\;,
\end{eqnarray}
where $Q$ denotes the conserved charge. These two operators act on the Wilson loop operator and the 't Hooft loop operator, respectively.
Thus, we are able to introduce two two-form background gauge fields of higher-form symmetries.

Inspired by this kind of higher-form symmetry realization, one can consider an axion-Maxwell theory coupled to a $\mathbb{Z}_n$ TQFT as shown in Ref.~\cite{Brennan:2023kpw}. The gauge field $A$ is an one-form gauge field and $F^{(2)}_A\equiv \partial\wedge A$ is its two-form field strength for a $U(1)_A$ theory which could be the $U(1)_{\rm EM}$ group in the SM. The action of the axion-Maxwell theory in this sector becomes
\begin{eqnarray}
S_0 = {1\over 2g^2}\int F^{(2)}_A F^{(2)}_A - {iK_A\over 8\pi^2 f_a} \int a F^{(2)}_A (F^{(2)}_A)^d\;,
\end{eqnarray}
where $f_a$ is the axion decay constant, and $K_A\in \mathbb{Z}$ is a discrete coupling constant. This axion-Maxwell theory is considered to couple to a $\mathbb{Z}_n$ gauge theory from a spontaneously broken $U(1)_B$ gauge theory. In this TQFT sector, the action of axion theory is
\begin{eqnarray}
S_1 = {in\over 2\pi}\int B^{(2)}(F^{(2)}_B)^d - {iK_B\over 4\pi^2 f_a} \int a F^{(2)}_B (F^{(2)}_B)^d - {iK_{AB}\over 8\pi^2 f_a} \int a F^{(2)}_A (F^{(2)}_B)^d\;,
\end{eqnarray}
where $F^{(2)}_B\equiv\partial \wedge B^{(1)}$ is the two-form field strength of an one-form $\mathbb{Z}_n$ gauge field $B^{(1)}$ associated with another $U(1)$ gauge group, and $B^{(2)}$ is a two-form gauge field associated with one-form $\mathbb{Z}_n^{(1)}$ gauge symmetry.
Then, the action of theory with topological QFT couplings via axion-portal can be given by $S_0+S_1$~\cite{Brennan:2023kpw}. The second term in $S_0$ and the last two terms in $S_1$ give more generic axion interactions.

It turns out that new axion interactions indeed arise based on different theories such as QEMD and the $\mathbb{Z}_n$ TQFT.
In this section we only review the existing framework in Ref.~\cite{Brennan:2023kpw} which realized new TQFT-couplings via axion-portal $aF_B^{(2)}(F^{(2)}_B)^d$ and $a F^{(2)}_A (F^{(2)}_B)^d$. Although they are analogous to those arisen based on QEMD theory, we are not trying to identify it to the QEMD theory. The $\mathbb{Z}_n$ gauge field $B^{(1)}$ discussed in Ref.~\cite{Brennan:2023kpw} and the four-potential $B_\mu$ of the Zwanziger theory arise in completely different theories. They have absolutely different kinetic terms and transform with respect to different gauge groups. The phenomenological studies of the TQFT theory require precise predictions in future work.

\section{Quantum calculation of axion-photon transition in QEMD}
\label{sec:Photon}

In this section we follow Ref.~\cite{Yang:2022uil} to perform the quantum calculations of axion-photon transition in QEMD under an external magnetic field or electric field as background.

\subsection{New axion-modified Maxwell equations}

Before performing the quantum calculations, we show the new axion-modified Maxwell equations in this framework.
The complete Lagrangian for the generic interactions between axion and four-potentials based on QEMD is~\cite{Sokolov:2022fvs}
\begin{eqnarray}
\mathcal{L}&=& {1\over 2n^2} \{[n\cdot (\partial \wedge B)]\cdot [n\cdot (\partial \wedge A)^d] -[n\cdot (\partial \wedge A)]\cdot [n\cdot (\partial \wedge B)^d] - [n\cdot (\partial\wedge A)]^2 \nonumber \\
&&- [n\cdot (\partial\wedge B)]^2\} -{1\over 4} g_{aAA}~a~{\rm tr}[(\partial \wedge A)(\partial \wedge A)^d] - {1\over 4} g_{aBB}~a~{\rm tr}[(\partial \wedge B)(\partial \wedge B)^d]\nonumber \\
&& - {1\over 2} g_{aAB}~a~{\rm tr}[(\partial \wedge A)(\partial \wedge B)^d] -j_e \cdot A - j_m\cdot B +\mathcal{L}_G \;,
\end{eqnarray}
where $\mathcal{L}_G$ is a gauge-fixing term.
The electromagnetic field strength tensor $F^{\mu\nu}$ and its dual tensor $F^{d~\mu\nu}$ are then introduced
\begin{eqnarray}
F=\partial \wedge A - (n\cdot \partial)^{-1} (n\wedge j_m)^d\;,~~
F^d=\partial \wedge B + (n\cdot \partial)^{-1} (n\wedge j_e)^d\;,
\label{eq:FFd}
\end{eqnarray}
where $j_e$ and $j_m$ are electric and magnetic currents, respectively.

After applying the Euler-Lagrange equation of motion for the two potentials, one obtains
\begin{eqnarray}
&&{1\over n^2}(n\cdot\partial n\cdot\partial A^\mu - n\cdot\partial \partial^\mu n\cdot A-n\cdot\partial n^\mu \partial\cdot A-n\cdot\partial \epsilon^\mu_{\nu\kappa\lambda}n^\nu \partial^\kappa B^\lambda) \nonumber \\
&&-g_{aAA}\partial_\nu a (\partial\wedge A)^{d~\nu\mu}-g_{aAB}\partial_\nu a (\partial\wedge B)^{d~\nu\mu}= j_e^\mu\;,\\
&&{1\over n^2}(n\cdot\partial n\cdot\partial B^\mu - n\cdot\partial \partial^\mu n\cdot B-n\cdot\partial n^\mu \partial\cdot B+n\cdot\partial \epsilon^\mu_{\nu\kappa\lambda}n^\nu \partial^\kappa A^\lambda) \nonumber \\
&&-g_{aBB}\partial_\nu a (\partial\wedge B)^{d~\nu\mu}-g_{aAB}\partial_\nu a (\partial\wedge A)^{d~\nu\mu}= j_m^\mu\;.
\end{eqnarray}
In terms of the field strength tensors $F^{\mu\nu}$ and $F^{d~\mu\nu}$, the above equations result in the following axion modified Maxwell equations~\cite{Sokolov:2022fvs}
\begin{eqnarray}
&&\partial_\mu F^{\mu\nu} -g_{aAA} \partial_\mu a F^{d~\mu\nu} + g_{aAB} \partial_\mu a F^{\mu\nu} = j_e^\nu\;,  \\
&&\partial_\mu F^{d~\mu\nu} +g_{aBB} \partial_\mu a F^{\mu\nu} - g_{aAB} \partial_\mu a F^{d~\mu\nu} = j_m^\nu\;,
\end{eqnarray}
where the term responsible for Witten effect is omitted.
The new Maxwell equations in terms of electric and magnetic fields are then given by
\begin{eqnarray}
&&\vec{\nabla}\times \vec{\mathbb{B}}-{\partial \vec{\mathbb{E}}\over \partial t}=\vec{j}_e+g_{aAA}(\vec{\mathbb{E}} \times \vec{\nabla} a - {\partial a\over \partial t} \vec{\mathbb{B}})
+ g_{aAB} (\vec{\mathbb{B}} \times \vec{\nabla} a + {\partial a\over \partial t} \vec{\mathbb{E}})\;,
\label{eq:Ampere}
\\
&&\vec{\nabla}\times \vec{\mathbb{E}}+{\partial \vec{\mathbb{B}}\over \partial t}=\vec{j}_m-g_{aBB}(\vec{\mathbb{B}} \times \vec{\nabla} a + {\partial a\over \partial t} \vec{\mathbb{E}})
- g_{aAB} (\vec{\mathbb{E}} \times \vec{\nabla} a - {\partial a\over \partial t} \vec{\mathbb{B}})\;,\\
&&\vec{\nabla}\cdot \vec{\mathbb{B}} = \rho_m -g_{aBB} \vec{\mathbb{E}}\cdot \vec{\nabla} a + g_{aAB} \vec{\mathbb{B}}\cdot \vec{\nabla} a \;,\\
&&\vec{\nabla}\cdot \vec{\mathbb{E}} = \rho_e + g_{aAA} \vec{\mathbb{B}}\cdot \vec{\nabla} a - g_{aAB} \vec{\mathbb{E}}\cdot \vec{\nabla} a \;,
\end{eqnarray}
where the magnetic charge $\rho_m$ and current $\vec{j}_m$ will be ignored below as there is no observed magnetic monopole.

\subsection{Magnetic background}

Suppose an external magnetic field $\mathbb{B}_0$ along the z-direction, according to Eq.~\eqref{eq:Lagrangian}, the axion and photon interaction can be written as
\begin{eqnarray}
\mathcal{L}_{a\gamma\gamma}=(g_{aAA}-g_{aBB})a\vec{\mathbb{E}}\cdot\Vec{\mathbb{B}}_0+g_{aAB}a\vec{\mathbb{B}}\cdot\Vec{\mathbb{B}}_0\;,
\label{eq:Lagrangian_H0}
\end{eqnarray}
where we have set external electric field to zero, i.e., $\vec{\mathbb{B}}_0=\hat{z}\mathbb{B}_0\neq 0$, $\vec{\mathbb{E}}_0=0$.
Due to the extremely light mass and low velocity of the axion DM, its de Brogile wavelength is of the order of $10^3/m_a$. It is much larger than the typical size of the cavity in haloscope experiments $\sim 1/m_a$.
Thus, the axion field inside the cavity can be approximately viewed as spatially independent and can be given in the form of cosine oscillation: $a(\vec{x},t)\approx a_0 \cos{\omega_a t}=\frac{\sqrt{2\rho_a}}{m_a}\cos{\omega_a t}$.
The Hamiltonian for the above interaction can be written as follows
\begin{eqnarray}
H_I=-\int{d^3x\mathcal{L}_{a\gamma\gamma}}=\frac{\sqrt{2\rho_a}}{m_a}\mathbb{B}_0\cos{(\omega_at)}\left[(g_{aBB}-g_{aAA})\int{d^3x \hat{z}\cdot \vec{\mathbb{E}}}-g_{aAB}\int{d^3x \hat{z}\cdot\vec{\mathbb{B}}}\right]\;.
\end{eqnarray}
We find that the key difference of axion in QEMD and QED lies in the axion induced electromagnetic fields. In QEMD, the axion induced fields $\vec{\mathbb{E}}$ and $\vec{\mathbb{B}}$ can occur simultaneously due to the presence of three couplings. As a result, the interactions between the axion field $a$ and the electromagnetic fields are divided into two parts: $g_{aBB}-g_{aAA}$ and $g_{aAB}$, corresponding to $\Vec{\mathbb{E}}$ and $\Vec{\mathbb{B}}$ respectively. However, in the traditional axion QED, only the $\hat{z}\cdot \vec{\mathbb{E}}$ term can appear given the substitution of $(g_{aBB}-g_{aAA})\to -g_{a\gamma\gamma}$.

Next, we will use a quantization approach to deal with the integrals $\int{d^3x \hat{z}\cdot\Vec{\mathbb{E}}}$ and $\int{d^3x\hat{z}\cdot\vec{\mathbb{B}}}$. In QEMD, the magnetic field and electric field can be given by using the curl of two vector potentials $\vec{A}$ and $\vec{B}$, respectively~\cite{Zwanziger:1970hk}
\begin{eqnarray}
-F^{d~0i}=\vec{\mathbb{B}}=\nabla\times\vec{A},~ -F^{0i}=\vec{\mathbb{E}}=-\nabla\times\vec{B}\;,
\label{eq:AB&HE}
\end{eqnarray}
in the absence of electric charge and magnetic charge.
We can expand the vector potentials $\vec{A}$ and $\vec{B}$ in terms of creation and annihilation operators as well as mode functions $\mathbf{u}_k(\mathbf{x})$
\begin{eqnarray}
    \vec{A}(\mathbf{x},t)=\sum_k\frac{1}{\sqrt{2\omega_kV}}(a_k \mathbf{u}_k^{(A)}(\bold x)e^{-i\omega_k t}+a_k^\dagger\bold{u}_k^{(A)*}(\bold{x})e^{i\omega_k t})\;,\\
    \vec{B}(\mathbf{x},t)=\sum_k\frac{1}{\sqrt{2\omega_kV}}(a_k \mathbf{u}_k^{(B)}(\bold x)e^{-i\omega_k t}+a_k^\dagger\bold{u}_k^{(B)*}(\bold{x})e^{i\omega_k t})\;,
\end{eqnarray}
where $V$ is the volume of a cavity, and $\mathbf{u}_k(\mathbf{x})$ functions satisfy the equations of motion with cavity-wall boundary conditions
\begin{eqnarray}
    \frac{n\cdot\partial}{n^2}(n\cdot\partial A^\mu-\partial^\mu n\cdot A-n^\mu\partial\cdot A-\epsilon^\mu_{\nu\rho\sigma}n^\nu \partial^\rho B^\sigma)=0\;,\\
    \frac{n\cdot\partial}{n^2}(n\cdot\partial B^\mu-\partial^\mu n\cdot B-n^\mu\partial\cdot B-\epsilon^\mu_{\nu\rho\sigma}n^\nu \partial^\rho A^\sigma)=0\;.
\end{eqnarray}
The operators $a_k$ and $a_k^\dagger$ can annihilate and create physical single-photon state even though two vector potentials are introduced to describe a photon. The above equations of motion are two first-order differential equations.
They constrain $A$ and $B$ together with the gauge-fixing condition
\begin{eqnarray}
\partial^2n\cdot A+\partial^2 n\cdot B=0\;.
\end{eqnarray}
They reduce the four degrees of $A$ and $B$ to the two degrees of freedom for a massless vector field. Furthermore, due to the equal-time commutation relations in Eq.~\eqref{eq:commutation1} and Eq.~\eqref{eq:commutation2}, the total degrees of freedom of photon can be further reduced to two.
Therefore, even if QEMD introduces two potentials $A$ and $B$, the degrees of freedom of physical photon are preserved.
Using the relations given by Eq.~\eqref{eq:AB&HE}, the electromagnetic fields can be obtained.
Although the exact forms of $\mathbf{u}_k^{(A,B)}$ are unknown, their curl in a cavity can be given by
\begin{eqnarray}
\nabla\times\mathbf{u}_k^{(A)}=\omega_k \mathbf{u}_k^\mathbb{B},~~~\nabla\times\mathbf{u}_k^{(B)}=\omega_k \mathbf{u}_k^\mathbb{E}\;,
\label{eq:relation1}
\end{eqnarray}
where $\mathbf{u}_k^\mathbb{B}$ and $\mathbf{u}_k^\mathbb{E}$ are the actual electromagnetic field modes inside the cavity with the normalization $\frac{1}{V}\int{d^3x\vert\mathbf{u}_k^{\mathbb{E},\mathbb{B}}\vert^2}=1$. Thus, we do not need to explicitly express the forms of $\mathbf{u}_k^{(A)}$ or $\mathbf{u}_k^{(B)}$ to do the following transition calculation but only show the results using $\mathbf{u}_k^\mathbb{B}$ and $\mathbf{u}_k^\mathbb{E}$.

One can then calculate the $\vert0\rangle\to \vert1\rangle$ photon transition matrix element as well as the transition probability inside the cavity with an external magnetic field $\vec{\mathbb{B}}_0$~\cite{Yang:2022uil}. Up to the first order, we have
\begin{eqnarray}
\begin{split}
P&\approx\left\vert\langle 1\vert \int_0^t dt H_I\vert0\rangle\right\vert^2\\
&\approx\frac{\rho_a}{m_a^2}\mathbb{B}_0^2V\Bigg[(g_{aBB}-g_{aAA})^2\sum_k \omega_k C_k^\mathbb{E}\\
&+g_{aAB}^2\sum_k \omega_k C_k^\mathbb{B}+2g_{aAB}(g_{aBB}-g_{aAA})\sum_k \omega_k C_k^{\mathbb{E}\mathbb{B}}\Bigg]\times\frac{\sin^2[(\omega_k-\omega_a)t/2]}{4[(\omega_k-\omega_a)/2]^2}\;,
\end{split}
\end{eqnarray}
where the relations in Eq.~\eqref{eq:relation1} are plugged into the above result, and $C_k^{\mathbb{E}, \mathbb{B}},~C_l^{\mathbb{E}\mathbb{B}}$ are the form factors that characterize the coupling strength of cavity mode $k$ to axions
\begin{eqnarray}
C_k^\mathbb{E}=\frac{\vert\int d^3x\hat{z}\cdot\mathbf{u}_k^\mathbb{E}\vert^2}{V\int{d^3x\vert \mathbf{u}_k^\mathbb{E}\vert^2}},~C_k^\mathbb{B}=\frac{\vert\int d^3x\hat{z}\cdot\mathbf{u}_k^\mathbb{B}\vert^2}{V\int{d^3x\vert \mathbf{u}_k^\mathbb{B}\vert^2}},~C_k^{\mathbb{E}\mathbb{B}}=\frac{\mathbf{Re}\left[\int d^3x\hat{z}\cdot\mathbf{u}_k^\mathbb{E} \int d^3x\hat{z}\cdot\mathbf{u}_k^{\mathbb{B}*}\right]}{V\sqrt{\int d^3x\vert\mathbf{u}_k^\mathbb{E}\vert^2\int d^3x\vert\mathbf{u}_k^\mathbb{B}\vert^2}}\;.
\label{eq:formfactor}
\end{eqnarray}
In practice, the transition emission process of a single photon is expected to take a long time $t$. The time factor can thus approximately become
\begin{eqnarray}
\frac{\sin^2[(\omega_k-\omega_a)t/2]}{4[(\omega_k-\omega_a)/2]^2}\approx \pi t\delta(\omega_k-\omega_a)/2\;.
\end{eqnarray}
Finally, the transition rate in the cavity can be obtained as
\begin{eqnarray}
R=dP/dt=\frac{\pi}{2}\frac{\rho_a}{m_a^2}\mathbb{B}_0^2VQ\left[(g_{aBB}-g_{aAA})^2C_{\omega_a}^\mathbb{E}+g_{aAB}^2 C_{\omega_a}^\mathbb{B}+2g_{aAB}(g_{aBB}-g_{aAA})C_{\omega_a}^{\mathbb{E}\mathbb{B}}\right],
\label{eq:rate}
\end{eqnarray}
where the discrete summation over the cavity modes $k$ is converted into continuous integrals with $\sum_k C_k^{\mathbb{E},\mathbb{B}} \omega_k \delta(\omega_k-\omega_a)\approx QC_{\omega_a}^{\mathbb{E},\mathbb{B}},~\sum_k C_k^{\mathbb{E}\mathbb{B}} \omega_k \delta(\omega_k-\omega_a)\approx QC_{\omega_a}^{\mathbb{E}\mathbb{B}}$. Note that the result of this quantum calculation under the assumption of $g_{aAA}\neq 0, g_{aBB}=g_{aAB}=0$ is larger than the classical result of the conventional axion cavity haloscope~\cite{Sikivie:1983ip,Krauss:1985ub} by a factor of $\pi/2$. We presume it is due to the underlying difference between classical physics and quantum physics.
For a given cavity in a haloscope experiment, $Q$ is assumed to be universal for any terms at the same moment $t$ and we can thus factor $Q$ out of the parentheses. This result clearly shows the enhancement of axion-photon transition by the cavity's quality factor when $\omega_k\approx \omega_a$.
The modes existing in a cylindrical cavity include TE modes (transverse electric modes with $\mathbb{E}_z=0,~\mathbb{B}_z\neq0$) and TM modes (transverse magnetic modes with $\mathbb{B}_z=0,~\mathbb{E}_z\neq0$), and there may be also some TEM modes ($\mathbb{E}_z=0,~\mathbb{B}_z=0$) embedded in.
Based on different types of cavity mode, the transition rate $R$ in Eq.~\eqref{eq:rate} can then be simplified as
\begin{eqnarray}
\begin{split}
R_{\text{TE}}&=\frac{\pi}{2}\frac{\rho_a}{m_a^2}\mathbb{B}_0^2VQg_{aAB}^2 C_{\omega_a}^\mathbb{B},~\text{with}~C^\mathbb{E}=C^{\mathbb{E}\mathbb{B}}=0\;,\\
R_{\text{TM}}&=\frac{\pi}{2}\frac{\rho_a}{m_a^2}\mathbb{B}_0^2VQ(g_{aBB}-g_{aAA})^2C_{\omega_a}^\mathbb{E},~\text{with}~C^\mathbb{B}=C^{\mathbb{E}\mathbb{B}}=0\;.
\label{eq:simplified rate}
\end{split}
\end{eqnarray}
In principle, which modes appear in the cavity depends on the choice of the direction of the external field.
For example, when the orientation of $\mathbb{B}_0$ is chosen to the z axis, only $\mathbb{B}_z$ induces axion signal for TE modes as we show above.
The similar conclusion also holds for TM modes.
Therefore, if we change the direction of the external field, the components other than $\mathbb{B}_z$ or $\mathbb{E}_z$ will appear.

Note that this approach has assumed that the final-state photon state is empty before the transition induced by axions. There is normally an ambient bath of thermal photons in a detector with a photon occupation number $n_\gamma$. One may think that the large occupation number could further boost the conversion rate by a factor of $n_\gamma+1$. However, based on the argument in Ref.~\cite{Ioannisian:2017srr}, there is always a backconversion of photons to axions in this case. As a result, the boost factor effect cancels out for the average photon production rate. Thus, with or without the ambient bath of thermal photons, the transition rate is the same.

In a cavity, the distribution of the electromagnetic field is usually quite different from that in vacuum. A cylindrical microwave resonant cavity can be viewed as a circular waveguide of length $L$ with short-circuit at both ends. Two movable bulk copper rods can be placed inside the cavity to achieve tuning frequency~\cite{ADMX:2021nhd}. The internal distribution of the electromagnetic field must satisfy both the Helmholtz equation and the corresponding boundary conditions, including those at the ends and on the walls of the cavity. The Helmholtz equation is
\begin{eqnarray}
\nabla^2u(r,\phi,z)+k^2u(r,\phi,z)=0,~\vec{\mathbb{E}}(r,\phi,z)~{\rm or}~\vec{\mathbb{B}}(r,\phi,z)=\hat{z}u(r,\phi,z)\;,
\end{eqnarray}
where only the $z$-component of the modes couples to the axion field according to the definition of the form factors in Eq.~\eqref{eq:formfactor}. Their solutions in different modes satisfy the following conditions
\begin{eqnarray}
\text{TE modes}~(\mathbb{E}_z=0,~\mathbb{B}_z=u(r,\phi,z)):
\left\{
\begin{array}{ll}
        r,~~\dfrac{\partial u(r,\phi,z)}{\partial r}\bigg\vert_{r=a}=0 \\
        \phi,~~u(r,\phi,z)=u(r,\phi+2\pi m,z)\\
        z,~~u(r,\phi,z)\bigg\vert_{z=0,~L}=0\;,
    \end{array}
\right. \\
    \text{TM modes}~(\mathbb{B}_z=0,~\mathbb{E}_z=u(r,\phi,z)):
\left\{
    \begin{array}{ll}
        r,~~u(r,\phi,z)\bigg\vert_{r=a}=0 \\
        \phi,~~u(r,\phi,z)=u(r,\phi+2\pi m,z)\\
        z,~~\dfrac{\partial u(r,\phi,z)}{\partial z}\bigg\vert_{z=0,~L}=0\;,
    \end{array}
\right.
\label{eq:Helmholtz}
\end{eqnarray}
where $a$ is the radius of circular cross section, $L$ is the length of cavity along the $z$-axis, and $m=(0,\pm 1,\pm 2,\cdots)$ represents a series of integers required by the periodic boundary conditions.
The solutions of the above differential equations yield a series of possible electromagnetic resonant modes that can exist inside the cavity
\begin{eqnarray}
\begin{split}
    \text{TE}_{mnp}:~\hat{z}\cdot \mathbf{u}_k^\mathbb{B}(r,\phi,z)=\mathbb{B}_z(r,\phi,z)_{mnp}&=\mathbb{B}_{mnp}J_m(k_\rho r)\left\{
    \begin{array}{ll}
        \cos{m\phi} \\
        \sin{m\phi}
    \end{array}
\right\}
\sin\left(\frac{p\pi z}{L} \right)\;,\\
\text{TM}_{mnp}:~\hat{z}\cdot \mathbf{u}_k^\mathbb{E}(r,\phi,z)=\mathbb{E}_z(r,\phi,z)_{mnp}&=\mathbb{E}_{mnp}J_m(k_\rho r)\left\{
    \begin{array}{ll}
        \cos{m\phi} \\
        \sin{m\phi}
    \end{array}
\right\}
\cos\left(\frac{p\pi z}{L} \right)\;,
\end{split}
\end{eqnarray}
where $\mathbb{B}_{mnp}$ and $\mathbb{E}_{mnp}$ are dimensionless coefficients ensuring the mode normalization.

\begin{figure}[htb!]
\centering
\includegraphics[scale=1,width=1\linewidth]{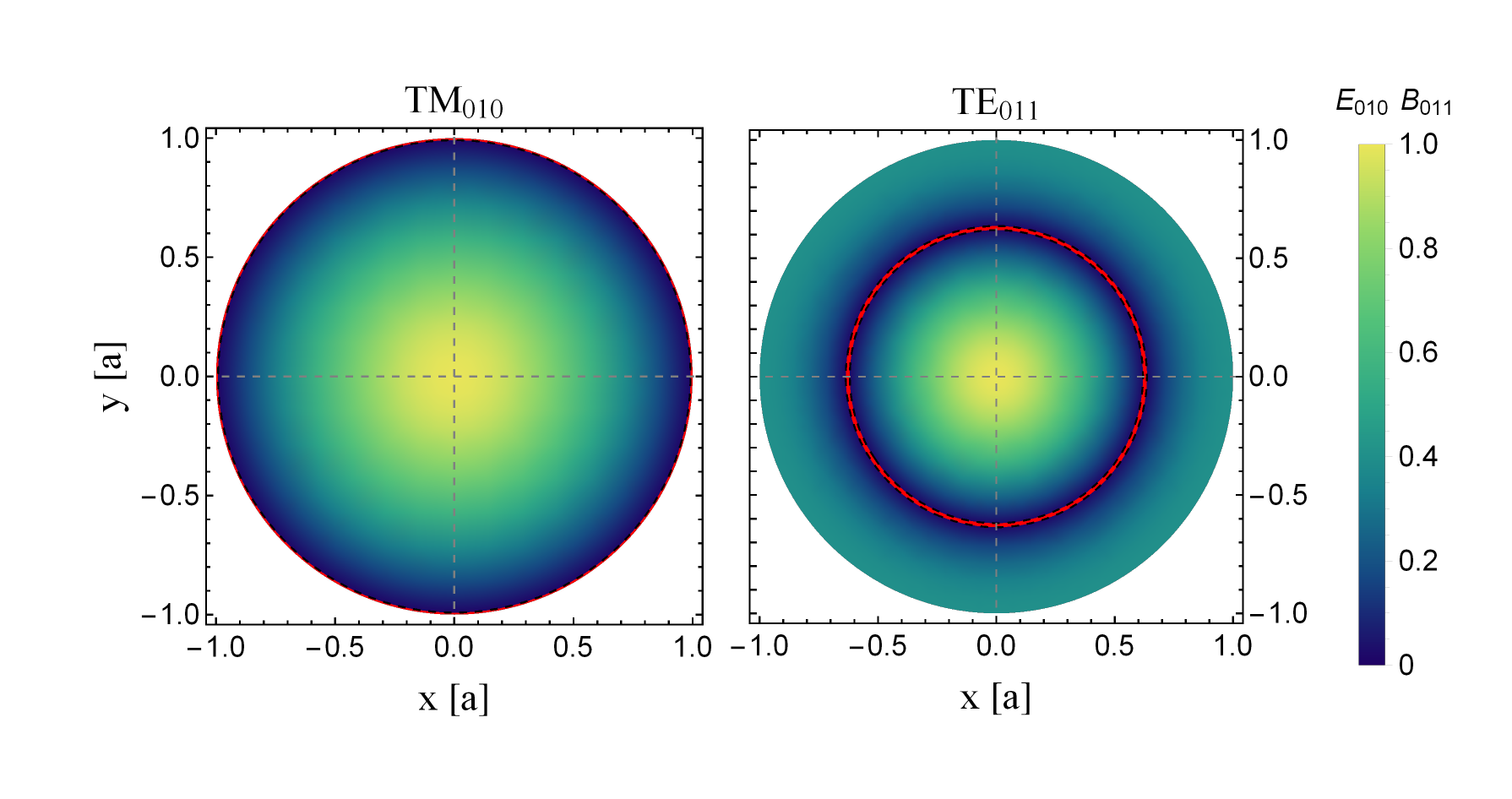}
\caption{Radial distributions of $\text{TM}_{010}$ (left) and $\text{TE}_{011}$ (right), where the transverse cross-section of TE mode  is taken at $z=L/2$. }
\label{fig:EH}
\end{figure}

From these solutions, it can be seen that the modes inside the cavity are represented by three integers $m$, $n$ and $p$ under the fixed boundary conditions. For TE modes, they are $m=(0,1,2,\cdots),~n=(1,2,3,\cdots),~p=(1,2,3,\cdots)$, and for TM modes $m=(0,1,2,\cdots),~n=(1,2,3,\cdots),~p=(0,1,2,\cdots)$. The eigenvalue of the radial part of Helmholtz equation is $k_\rho=(x_{m,n}^\prime/a)$ for TE modes or $k_\rho=(x_{m,n}/a)$ for TM modes with $x$ and $x^\prime$ being the $n$-th zero points of the Bessel function $J_m(x)$ and its first derivative $J^\prime_m(x)$, respectively.
In this way, the form factor can be obtained by plugging the above solutions into their definitions in Eq.~(\ref{eq:formfactor}) and integrating over the volume of the cavity. For the TM modes, only when $m=p=0$, the integrals in the $z$ and $\phi$ directions are non-zero. Therefore, we only consider the $\text{TM}_{010}$ mode. For the TE modes, unlike TM modes, they satisfy the second boundary condition in $r$ direction and the derivative of field is zero on the cavity wall. This leads to its vanishing integral over the cavity wall, even if those in $z$ and $\phi$ directions are non-zero ($m=0,~p=1,3,5,\cdots$). Fig.~\ref{fig:EH} shows the radial distributions of two modes, $\text{TM}_{010}$ and $\text{TE}_{011}$. For the $\text{TM}_{010}$ mode, the amplitude of field strength $\mathbb{E}_{010}$ decreases from the maximum at the cavity center to zero on the cavity wall. The field strength $\mathbb{B}_{011}$ decreases to zero at the red circle and instead increases outside the circle, resulting in the cancellation of the cavity response to the axion.

Thus, one can conclude that in a cylindrical cavity TE mode has no coupling with axion, i.e $C^{\mathbb{B}}=0$, and $R_{\text{TE}}$ in Eq.~\eqref{eq:simplified rate} is zero. It means under an external magnetic field $\mathbb{B}_0$, only the coupling $(g_{aBB}-g_{aAA})\approx g_{aBB}$ can be measured through the TM mode. For illustration, we show the transition rate $R_{\text{TM}}$ in terms of practical units as
\begin{eqnarray}
R_{\text{TM}}\approx3.63~\text{Hz}\left(\frac{\rho_a}{7.1\times 10^{-25}~\text{g}/\text{cm}^3}\right)\left(\frac{10^{-5}~\text{eV}}{m_a}\right)^2\left(\frac{\mathbb{B}_0}{10~\text{mT}}\right)^2\nonumber\\
\cdot\left(\frac{V}{0.001~\text{m}^3}\right)\left(\frac{Q}{10^5}\right)\left(\frac{g_{aBB}}{10^{-12}~\text{GeV}^{-1}}\right)^2\left(\frac{C^\mathbb{E}_{\omega_a}}{1}\right)\;,
\label{eq:R_TM_units}
\end{eqnarray}
where the density $7.1\times 10^{-25}~\text{g}/\text{cm}^3$ corresponds to the dark matter local density $0.4~\rm{GeV}/cm^3$~\cite{Workman:2022ynf}.
Given this parameter setup, the axion cavity can be viewed as a device that emits single photon with a slow rate. The resolution of existing linear detectors is large enough to detect this signal. This approach is exactly the same as that measuring the conventional axion coupling $g_{a\gamma\gamma}$ in axion electrodynamics.

\subsection{Electric background}

When the external magnetic field is replaced by an external electric field $\mathbb{E}_0$, Eq.~\eqref{eq:Lagrangian_H0} can be rewritten as
\begin{eqnarray}
\mathcal{L}_{a\gamma\gamma}=(g_{aAA}-g_{aBB})a\vec{\mathbb{B}}\cdot\vec{\mathbb{E}}_0-g_{aAB}a\vec{\mathbb{E}}\cdot\vec{\mathbb{E}}_0\;.
\end{eqnarray}
Similarly, the photon emission rates for TE and TM modes in the axion cavity can be obtained as
\begin{eqnarray}
\begin{split}
R_{\text{TE}}&=\frac{\pi}{2}\frac{\rho_a}{m_a^2}\mathbb{E}_0^2VQ(g_{aBB}-g_{aAA})^2C_{\omega_a}^\mathbb{B},~\text{with}~C^\mathbb{E}=C^{\mathbb{E}\mathbb{B}}=0\;,\\
R_{\text{TM}}&=\dfrac{\pi}{2}\dfrac{\rho_a}{m_a^2}\mathbb{E}_0^2VQg_{aAB}^2C_{\omega_a}^\mathbb{E},~\text{with}~C^\mathbb{B}=C^{\mathbb{E}\mathbb{B}}=0\;.
\label{eq:simplified rate_E0}
\end{split}
\end{eqnarray}
We find that replacing $\mathbb{B}_0$ with $\mathbb{E}_0$ is associated with exchanging the couplings constrained by TE and TM modes. Now, the sensitivity to $g_{aAB}$ is still given by the transverse magnetic wave which couples to axion.
Similar to Eq.~\eqref{eq:R_TM_units}, the transition rate $R_{\text{TM}}$ under the external electric field is given by
\begin{eqnarray}
R_{\text{TM}}=0.41~\text{Hz}\left(\frac{\rho_a}{7.1\times 10^{-25}~\text{g}/\text{cm}^3}\right)\left(\frac{10^{-5}~\text{eV}}{m_a}\right)^2\left(\frac{\mathbb{E}_0}{10^3~\text{kV/m}}\right)^2\nonumber\\
\cdot\left(\frac{V}{0.001~\text{m}^3}\right)\left(\frac{Q}{10^5}\right)\left(\frac{g_{aAB}}{10^{-12}~\text{GeV}^{-1}}\right)^2\left(\frac{C^\mathbb{E}_{\omega_a}}{1}\right)\;.
\label{eq:practical_unit2}
\end{eqnarray}

\section{Sensitivity of resonant cavity to axion-photon couplings in QEMD}
\label{sec:Cavity}

Based on the above arguments, we propose the following scheme for detecting axion interactions in QEMD through the cavity haloscope. One can apply an external magnetic field $\mathbb{B}_0$ to measure $(g_{aBB}-g_{aAA})\approx g_{aBB}$ coupling or an external electric field $\mathbb{E}_0$ to measure $g_{aAB}$ coupling. The corresponding sensitivities are given by the experimental configuration and the form factor $C^\mathbb{E}_{\omega_a}$ in TM mode.

For the $\text{TM}_{0n0}$ mode, the form factor $C^\mathbb{E}_{\omega_a}$ has a simple analytic result: $C^\mathbb{E}_{\omega_a}=C^\mathbb{E}_{0n0}=4/(x_{0,n})^2$. Once the mode is fixed, the value of the form factor can be determined and is independent of $a$ and $L$. For $\text{TM}_{010}$, one gets $C^{\mathbb{E}}_{010}=0.69$~\cite{Sikivie:2020zpn}. Note that the axion energy is determined by the eigenvalues $k_\rho$ and $k_z$ of the Helmholtz equation in Eq.~\eqref{eq:Helmholtz}, that is, $\omega_a=k=\sqrt{k_\rho^2+k_z^2}$ which simplifies to $\omega_a=(x_{0,1})/a$ for the $\text{TM}_{010}$ mode with $x_{0,1}=2.4$. It means that the axion mass corresponding to the experimentally measured axion coupling is only related to the radius $a$ of the cavity. When $a$ is 5 cm, one has $\omega_a\approx 9.5\times 10^{-6}$ eV. To determine the sensitivity in other mass regions, tuning is required by changing the value of $a$. In practical experiments such as ADMX~\cite{ADMX:2021nhd}, two movable bulk copper rods are placed inside the cavity to achieve tuning. Here, we do not intend to explore the details of techniques for adjusting the resonant frequency by changing the cavity structure, but only provide the cavity radius $a$ required for the resonance condition when the axion mass is $m_a=\omega_a$.

\begin{figure}
\centering
\includegraphics[scale=1,width=0.8\linewidth]{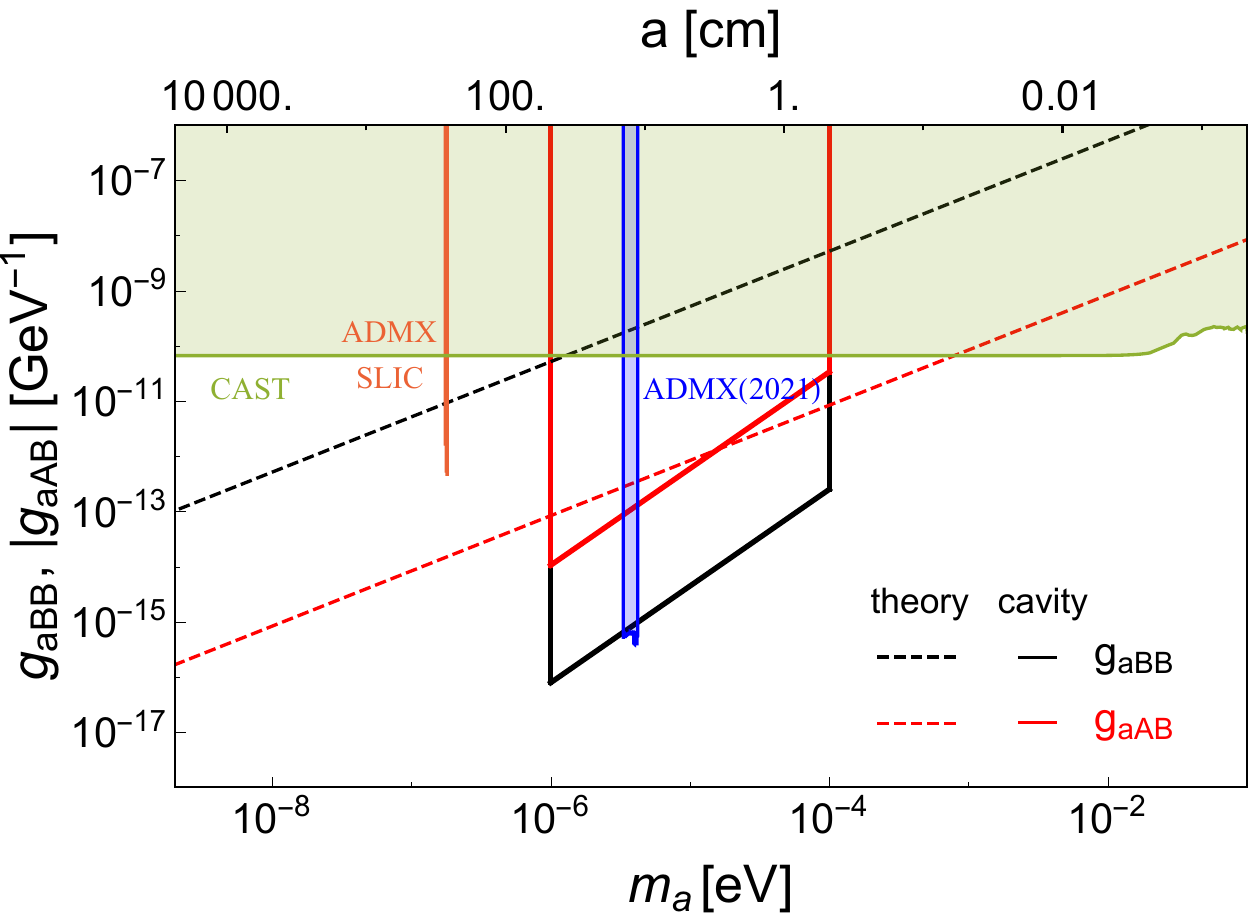}
\caption{The expected sensitivity of $|g_{aAB}|$ (red solid line) and $g_{aBB}$ (black solid line) in the cavity haloscope experiment. The dashed lines indicate the corresponding theoretical predictions (red for $|g_{aAB}|$, black for $g_{aBB}$). Existing limits from ADMX (2021)~\cite{ADMX:2021nhd}, ADMX SLIC~\cite{Crisosto:2019fcj} and CAST~\cite{CAST:2017uph} are also shown for comparison.}
\label{fig:sensitivity}
\end{figure}

In an external magnetic field, the signal power is given by
\begin{eqnarray}
P_{\text{signal}}=m_aR_{\text{TM}}=\frac{\pi}{2}\frac{\rho_a}{m_a}\mathbb{B}_0^2VQg_{aBB}^2C^\mathbb{E}_{010}\;,
\end{eqnarray}
where the cavity volume $V$ is regarded as a function of $m_a$, i.e. $V=\pi a^2 L=\pi L (x_{0,1}/m_a)^2$, to ensure that $m_a$ corresponding to each sensitivity is always at the resonant point. Similarly, the signal power under external electric field can be obtained by making a replacement $g_{aBB}\mathbb{B}_0 \rightarrow{g_{aAB}\mathbb{E}_0}$. A typical detection device of axion cavity consists of a main cavity and an amplification chain. The main source of noise comes from the cryogenic high electron mobility transistor (HEMT). It contributes to an effective noise temperature $T_{\text{eff}}$ around a few Kelvins, but it will be further enhanced to above 10 K due to the microwave loss in channel as the microwave signal is emitted from cavity and then received by HEMT~\cite{Yang:2022uil}.
Currently, many axion cavity experiments have applied Josephson Parametric Amplifiers (JPA) to enhance detection sensitivity. For example, in ADMX, one can control the noise temperature at the order of $\mathcal{O}(10^2)$ mK~\cite{ADMX:2021nhd}.
Therefore, the signal-to-noise ratio is given by
\begin{eqnarray}
\text{SNR}=\frac{P_{\text{signal}}}{k_BT_{\text{eff}}}\sqrt{\frac{t}{b}}\;,
\end{eqnarray}
where $k_B$ is the Boltzmann constant, $t$ is the observation time and the ratio of frequency and $Q$ factor is the detector bandwidth $b=f/Q$. To estimate the sensitivity of cavity experiment to $g_{aBB}$ or $g_{aAB}$, we take $Q=10^5$ and limit SNR to 5. The results of sensitivity bound are shown in Fig.~\ref{fig:sensitivity}. Assuming an observation time of 90 seconds~\cite{Hagmann:1990tj} and a cavity length of 1 m, we obtain the corresponding bounds on couplings $g_{aBB}$ and $g_{aAB}$ in the external magnetic field $\mathbb{B}_0=10$ T and electric field $\mathbb{E}_0=10^4$ kV/m, respectively.
For $g_{aBB}$, the temperature parameter inside the cavity is assumed to be $T_{\text{eff}}=0.5$ K, and for $g_{aAB}$ it is $T_{\text{eff}}=0.1$ K. It should be noticed that $m_a$ cannot be arbitrarily small or large since the smaller $m_a$ is, the larger the cavity radius $a$ is required to satisfy the resonance condition, and vice versa for larger mass of axion. On the other hand, when $m_a$ is too small, the transition rate $R$ increases sharply to GHz level and exceeds the detector resolution. Thus, we only consider axions within the mass range of $10^{-6}\sim 10^{-4}$~eV.
It turns out that the theoretical predictions of new axion couplings can be probed in this mass range.
The result of current cavity experiments such as ADMX measuring the conventional axion coupling $g_{a\gamma\gamma}$ can apply to confine the $g_{aBB}$ coupling. The existing ADMX bound given $\mathbb{B}_0=7.5$ T and $T_{\text{eff}}=0.5$ K has already excluded a part of parameter space of $g_{aBB}$ coupling. To measure $g_{aAB}$, a strong electric field $\mathbb{E}_0$ and lower temperature are both required. The same constraint of helioscope search also holds for the new $g_{aBB}$ coupling. In Fig.~\ref{fig:sensitivity}, we add CAST constraint~\cite{CAST:2017uph} for reference. It excludes the theoretically predicted $g_{aBB}$ coupling in the axion mass range of $10^{-6}\sim 10^{-4}$~eV.

\section{Conclusion}
\label{sec:Con}

Motivated by the Witten effect, the axion-dyon dynamics can be reliably built based on the quantum electromagnetodynamics.
Two $U(1)$ gauge groups and two four-potentials $A^\mu$ and $B^\mu$ are introduced to describe electric charge, magnetic charge and photon in this framework. As a result, three anomalous interactions between axion and photon arise in contrary to the conventional axion-photon coupling $g_{a\gamma\gamma}$.
We also review a generic low-energy axion-photon effective field theory can also be realized in the language of ``generalized symmetries'' with higher-form symmetries and gauge fields based on a $\mathbb{Z}_n$ TQFT.

In this work, we provide a complete quantum calculation of axion-single photon transition rate inside a homogeneous electromagnetic field in terms of the new axion interaction Hamiltonian in QEMD. This quantum calculation can clearly imply the enhancement of conversion rate through resonant cavity in axion haloscope experiments. Our work provides the basic method for the generic cavity search of new axion-photon couplings in QEMD framework. We find that an external magnetic field $\mathbb{B}_0$ can be set to measure $(g_{aBB}-g_{aAA})\approx g_{aBB}$ coupling or an external electric field $\mathbb{E}_0$ to measure $g_{aAB}$ coupling. The corresponding sensitivity bounds are given by the cavity experimental configuration and the form factor $C^\mathbb{E}_{010}$ in TM mode for the QEMD axion couplings.

\acknowledgments
We thank Anton V.~Sokolov, Andreas Ringwald, Yu Gao and Qiaoli Yang for useful comment and discussion. T.~L. is supported by the National Natural Science Foundation of China (Grant No. 12375096, 11975129, 12035008) and ``the Fundamental Research Funds for the Central Universities'', Nankai University (Grants No. 63196013).

\bibliography{refs}

\begin{thebibliography}{51}%
\makeatletter
\providecommand \@ifxundefined [1]{%
 \@ifx{#1\undefined}
}%
\providecommand \@ifnum [1]{%
 \ifnum #1\expandafter \@firstoftwo
 \else \expandafter \@secondoftwo
 \fi
}%
\providecommand \@ifx [1]{%
 \ifx #1\expandafter \@firstoftwo
 \else \expandafter \@secondoftwo
 \fi
}%
\providecommand \natexlab [1]{#1}%
\providecommand \enquote  [1]{``#1''}%
\providecommand \bibnamefont  [1]{#1}%
\providecommand \bibfnamefont [1]{#1}%
\providecommand \citenamefont [1]{#1}%
\providecommand \href@noop [0]{\@secondoftwo}%
\providecommand \href [0]{\begingroup \@sanitize@url \@href}%
\providecommand \@href[1]{\@@startlink{#1}\@@href}%
\providecommand \@@href[1]{\endgroup#1\@@endlink}%
\providecommand \@sanitize@url [0]{\catcode `\\12\catcode `\$12\catcode
  `\&12\catcode `\#12\catcode `\^12\catcode `\_12\catcode `\%12\relax}%
\providecommand \@@startlink[1]{}%
\providecommand \@@endlink[0]{}%
\providecommand \url  [0]{\begingroup\@sanitize@url \@url }%
\providecommand \@url [1]{\endgroup\@href {#1}{\urlprefix }}%
\providecommand \urlprefix  [0]{URL }%
\providecommand \Eprint [0]{\href }%
\providecommand \doibase [0]{http://dx.doi.org/}%
\providecommand \selectlanguage [0]{\@gobble}%
\providecommand \bibinfo  [0]{\@secondoftwo}%
\providecommand \bibfield  [0]{\@secondoftwo}%
\providecommand \translation [1]{[#1]}%
\providecommand \BibitemOpen [0]{}%
\providecommand \bibitemStop [0]{}%
\providecommand \bibitemNoStop [0]{.\EOS\space}%
\providecommand \EOS [0]{\spacefactor3000\relax}%
\providecommand \BibitemShut  [1]{\csname bibitem#1\endcsname}%
\let\auto@bib@innerbib\@empty
\bibitem [{\citenamefont {Peccei}\ and\ \citenamefont
  {Quinn}(1977{\natexlab{a}})}]{Peccei:1977hh}%
  \BibitemOpen
  \bibfield  {author} {\bibinfo {author} {\bibfnamefont {R.~D.}\ \bibnamefont
  {Peccei}}\ and\ \bibinfo {author} {\bibfnamefont {H.~R.}\ \bibnamefont
  {Quinn}},\ }\href {\doibase 10.1103/PhysRevLett.38.1440} {\bibfield
  {journal} {\bibinfo  {journal} {Phys. Rev. Lett.}\ }\textbf {\bibinfo
  {volume} {38}},\ \bibinfo {pages} {1440} (\bibinfo {year}
  {1977}{\natexlab{a}})}\BibitemShut {NoStop}%
\bibitem [{\citenamefont {Peccei}\ and\ \citenamefont
  {Quinn}(1977{\natexlab{b}})}]{Peccei:1977ur}%
  \BibitemOpen
  \bibfield  {author} {\bibinfo {author} {\bibfnamefont {R.~D.}\ \bibnamefont
  {Peccei}}\ and\ \bibinfo {author} {\bibfnamefont {H.~R.}\ \bibnamefont
  {Quinn}},\ }\href {\doibase 10.1103/PhysRevD.16.1791} {\bibfield  {journal}
  {\bibinfo  {journal} {Phys. Rev. D}\ }\textbf {\bibinfo {volume} {16}},\
  \bibinfo {pages} {1791} (\bibinfo {year} {1977}{\natexlab{b}})}\BibitemShut
  {NoStop}%
\bibitem [{\citenamefont {Weinberg}(1978)}]{Weinberg:1977ma}%
  \BibitemOpen
  \bibfield  {author} {\bibinfo {author} {\bibfnamefont {S.}~\bibnamefont
  {Weinberg}},\ }\href {\doibase 10.1103/PhysRevLett.40.223} {\bibfield
  {journal} {\bibinfo  {journal} {Phys. Rev. Lett.}\ }\textbf {\bibinfo
  {volume} {40}},\ \bibinfo {pages} {223} (\bibinfo {year} {1978})}\BibitemShut
  {NoStop}%
\bibitem [{\citenamefont {Wilczek}(1978)}]{Wilczek:1977pj}%
  \BibitemOpen
  \bibfield  {author} {\bibinfo {author} {\bibfnamefont {F.}~\bibnamefont
  {Wilczek}},\ }\href {\doibase 10.1103/PhysRevLett.40.279} {\bibfield
  {journal} {\bibinfo  {journal} {Phys. Rev. Lett.}\ }\textbf {\bibinfo
  {volume} {40}},\ \bibinfo {pages} {279} (\bibinfo {year} {1978})}\BibitemShut
  {NoStop}%
\bibitem [{\citenamefont {Baluni}(1979)}]{Baluni:1978rf}%
  \BibitemOpen
  \bibfield  {author} {\bibinfo {author} {\bibfnamefont {V.}~\bibnamefont
  {Baluni}},\ }\href {\doibase 10.1103/PhysRevD.19.2227} {\bibfield  {journal}
  {\bibinfo  {journal} {Phys. Rev. D}\ }\textbf {\bibinfo {volume} {19}},\
  \bibinfo {pages} {2227} (\bibinfo {year} {1979})}\BibitemShut {NoStop}%
\bibitem [{\citenamefont {Crewther}\ \emph {et~al.}(1979)\citenamefont
  {Crewther}, \citenamefont {Di~Vecchia}, \citenamefont {Veneziano},\ and\
  \citenamefont {Witten}}]{Crewther:1979pi}%
  \BibitemOpen
  \bibfield  {author} {\bibinfo {author} {\bibfnamefont {R.~J.}\ \bibnamefont
  {Crewther}}, \bibinfo {author} {\bibfnamefont {P.}~\bibnamefont
  {Di~Vecchia}}, \bibinfo {author} {\bibfnamefont {G.}~\bibnamefont
  {Veneziano}}, \ and\ \bibinfo {author} {\bibfnamefont {E.}~\bibnamefont
  {Witten}},\ }\href {\doibase 10.1016/0370-2693(79)90128-X} {\bibfield
  {journal} {\bibinfo  {journal} {Phys. Lett. B}\ }\textbf {\bibinfo {volume}
  {88}},\ \bibinfo {pages} {123} (\bibinfo {year} {1979})},\ \bibinfo {note}
  {[Erratum: Phys.Lett.B 91, 487 (1980)]}\BibitemShut {NoStop}%
\bibitem [{\citenamefont {Kim}(1979)}]{Kim:1979if}%
  \BibitemOpen
  \bibfield  {author} {\bibinfo {author} {\bibfnamefont {J.~E.}\ \bibnamefont
  {Kim}},\ }\href {\doibase 10.1103/PhysRevLett.43.103} {\bibfield  {journal}
  {\bibinfo  {journal} {Phys. Rev. Lett.}\ }\textbf {\bibinfo {volume} {43}},\
  \bibinfo {pages} {103} (\bibinfo {year} {1979})}\BibitemShut {NoStop}%
\bibitem [{\citenamefont {Shifman}\ \emph {et~al.}(1980)\citenamefont
  {Shifman}, \citenamefont {Vainshtein},\ and\ \citenamefont
  {Zakharov}}]{Shifman:1979if}%
  \BibitemOpen
  \bibfield  {author} {\bibinfo {author} {\bibfnamefont {M.~A.}\ \bibnamefont
  {Shifman}}, \bibinfo {author} {\bibfnamefont {A.~I.}\ \bibnamefont
  {Vainshtein}}, \ and\ \bibinfo {author} {\bibfnamefont {V.~I.}\ \bibnamefont
  {Zakharov}},\ }\href {\doibase 10.1016/0550-3213(80)90209-6} {\bibfield
  {journal} {\bibinfo  {journal} {Nucl. Phys. B}\ }\textbf {\bibinfo {volume}
  {166}},\ \bibinfo {pages} {493} (\bibinfo {year} {1980})}\BibitemShut
  {NoStop}%
\bibitem [{\citenamefont {Dine}\ \emph {et~al.}(1981)\citenamefont {Dine},
  \citenamefont {Fischler},\ and\ \citenamefont {Srednicki}}]{Dine:1981rt}%
  \BibitemOpen
  \bibfield  {author} {\bibinfo {author} {\bibfnamefont {M.}~\bibnamefont
  {Dine}}, \bibinfo {author} {\bibfnamefont {W.}~\bibnamefont {Fischler}}, \
  and\ \bibinfo {author} {\bibfnamefont {M.}~\bibnamefont {Srednicki}},\ }\href
  {\doibase 10.1016/0370-2693(81)90590-6} {\bibfield  {journal} {\bibinfo
  {journal} {Phys. Lett. B}\ }\textbf {\bibinfo {volume} {104}},\ \bibinfo
  {pages} {199} (\bibinfo {year} {1981})}\BibitemShut {NoStop}%
\bibitem [{\citenamefont {Zhitnitsky}(1980)}]{Zhitnitsky:1980tq}%
  \BibitemOpen
  \bibfield  {author} {\bibinfo {author} {\bibfnamefont {A.~R.}\ \bibnamefont
  {Zhitnitsky}},\ }\href@noop {} {\bibfield  {journal} {\bibinfo  {journal}
  {Sov. J. Nucl. Phys.}\ }\textbf {\bibinfo {volume} {31}},\ \bibinfo {pages}
  {260} (\bibinfo {year} {1980})}\BibitemShut {NoStop}%
\bibitem [{\citenamefont {Baker}\ \emph {et~al.}(2006)\citenamefont {Baker}
  \emph {et~al.}}]{Baker:2006ts}%
  \BibitemOpen
  \bibfield  {author} {\bibinfo {author} {\bibfnamefont {C.~A.}\ \bibnamefont
  {Baker}} \emph {et~al.},\ }\href {\doibase 10.1103/PhysRevLett.97.131801}
  {\bibfield  {journal} {\bibinfo  {journal} {Phys. Rev. Lett.}\ }\textbf
  {\bibinfo {volume} {97}},\ \bibinfo {pages} {131801} (\bibinfo {year}
  {2006})},\ \Eprint {http://arxiv.org/abs/hep-ex/0602020}
  {arXiv:hep-ex/0602020} \BibitemShut {NoStop}%
\bibitem [{\citenamefont {Pendlebury}\ \emph {et~al.}(2015)\citenamefont
  {Pendlebury} \emph {et~al.}}]{Pendlebury:2015lrz}%
  \BibitemOpen
  \bibfield  {author} {\bibinfo {author} {\bibfnamefont {J.~M.}\ \bibnamefont
  {Pendlebury}} \emph {et~al.},\ }\href {\doibase 10.1103/PhysRevD.92.092003}
  {\bibfield  {journal} {\bibinfo  {journal} {Phys. Rev. D}\ }\textbf {\bibinfo
  {volume} {92}},\ \bibinfo {pages} {092003} (\bibinfo {year} {2015})},\
  \Eprint {http://arxiv.org/abs/1509.04411} {arXiv:1509.04411 [hep-ex]}
  \BibitemShut {NoStop}%
\bibitem [{\citenamefont {Witten}(1979)}]{Witten:1979ey}%
  \BibitemOpen
  \bibfield  {author} {\bibinfo {author} {\bibfnamefont {E.}~\bibnamefont
  {Witten}},\ }\href {\doibase 10.1016/0370-2693(79)90838-4} {\bibfield
  {journal} {\bibinfo  {journal} {Phys. Lett. B}\ }\textbf {\bibinfo {volume}
  {86}},\ \bibinfo {pages} {283} (\bibinfo {year} {1979})}\BibitemShut
  {NoStop}%
\bibitem [{\citenamefont {Fischler}\ and\ \citenamefont
  {Preskill}(1983)}]{Fischler:1983sc}%
  \BibitemOpen
  \bibfield  {author} {\bibinfo {author} {\bibfnamefont {W.}~\bibnamefont
  {Fischler}}\ and\ \bibinfo {author} {\bibfnamefont {J.}~\bibnamefont
  {Preskill}},\ }\href {\doibase 10.1016/0370-2693(83)91260-1} {\bibfield
  {journal} {\bibinfo  {journal} {Phys. Lett. B}\ }\textbf {\bibinfo {volume}
  {125}},\ \bibinfo {pages} {165} (\bibinfo {year} {1983})}\BibitemShut
  {NoStop}%
\bibitem [{\citenamefont {Kawasaki}\ \emph {et~al.}(2016)\citenamefont
  {Kawasaki}, \citenamefont {Takahashi},\ and\ \citenamefont
  {Yamada}}]{Kawasaki:2015lpf}%
  \BibitemOpen
  \bibfield  {author} {\bibinfo {author} {\bibfnamefont {M.}~\bibnamefont
  {Kawasaki}}, \bibinfo {author} {\bibfnamefont {F.}~\bibnamefont {Takahashi}},
  \ and\ \bibinfo {author} {\bibfnamefont {M.}~\bibnamefont {Yamada}},\ }\href
  {\doibase 10.1016/j.physletb.2015.12.075} {\bibfield  {journal} {\bibinfo
  {journal} {Phys. Lett. B}\ }\textbf {\bibinfo {volume} {753}},\ \bibinfo
  {pages} {677} (\bibinfo {year} {2016})},\ \Eprint
  {http://arxiv.org/abs/1511.05030} {arXiv:1511.05030 [hep-ph]} \BibitemShut
  {NoStop}%
\bibitem [{\citenamefont {Nomura}\ \emph {et~al.}(2016)\citenamefont {Nomura},
  \citenamefont {Rajendran},\ and\ \citenamefont {Sanches}}]{Nomura:2015xil}%
  \BibitemOpen
  \bibfield  {author} {\bibinfo {author} {\bibfnamefont {Y.}~\bibnamefont
  {Nomura}}, \bibinfo {author} {\bibfnamefont {S.}~\bibnamefont {Rajendran}}, \
  and\ \bibinfo {author} {\bibfnamefont {F.}~\bibnamefont {Sanches}},\ }\href
  {\doibase 10.1103/PhysRevLett.116.141803} {\bibfield  {journal} {\bibinfo
  {journal} {Phys. Rev. Lett.}\ }\textbf {\bibinfo {volume} {116}},\ \bibinfo
  {pages} {141803} (\bibinfo {year} {2016})},\ \Eprint
  {http://arxiv.org/abs/1511.06347} {arXiv:1511.06347 [hep-ph]} \BibitemShut
  {NoStop}%
\bibitem [{\citenamefont {Kawasaki}\ \emph {et~al.}(2018)\citenamefont
  {Kawasaki}, \citenamefont {Takahashi},\ and\ \citenamefont
  {Yamada}}]{Kawasaki:2017xwt}%
  \BibitemOpen
  \bibfield  {author} {\bibinfo {author} {\bibfnamefont {M.}~\bibnamefont
  {Kawasaki}}, \bibinfo {author} {\bibfnamefont {F.}~\bibnamefont {Takahashi}},
  \ and\ \bibinfo {author} {\bibfnamefont {M.}~\bibnamefont {Yamada}},\ }\href
  {\doibase 10.1007/JHEP01(2018)053} {\bibfield  {journal} {\bibinfo  {journal}
  {JHEP}\ }\textbf {\bibinfo {volume} {01}},\ \bibinfo {pages} {053} (\bibinfo
  {year} {2018})},\ \Eprint {http://arxiv.org/abs/1708.06047} {arXiv:1708.06047
  [hep-ph]} \BibitemShut {NoStop}%
\bibitem [{\citenamefont {Houston}\ and\ \citenamefont
  {Li}(2017)}]{Houston:2017kwe}%
  \BibitemOpen
  \bibfield  {author} {\bibinfo {author} {\bibfnamefont {N.}~\bibnamefont
  {Houston}}\ and\ \bibinfo {author} {\bibfnamefont {T.}~\bibnamefont {Li}},\
  }\href@noop {} {\  (\bibinfo {year} {2017})},\ \Eprint
  {http://arxiv.org/abs/1711.05721} {arXiv:1711.05721 [hep-ph]} \BibitemShut
  {NoStop}%
\bibitem [{\citenamefont {Sato}\ \emph {et~al.}(2018)\citenamefont {Sato},
  \citenamefont {Takahashi},\ and\ \citenamefont {Yamada}}]{Sato:2018nqy}%
  \BibitemOpen
  \bibfield  {author} {\bibinfo {author} {\bibfnamefont {R.}~\bibnamefont
  {Sato}}, \bibinfo {author} {\bibfnamefont {F.}~\bibnamefont {Takahashi}}, \
  and\ \bibinfo {author} {\bibfnamefont {M.}~\bibnamefont {Yamada}},\ }\href
  {\doibase 10.1103/PhysRevD.98.043535} {\bibfield  {journal} {\bibinfo
  {journal} {Phys. Rev. D}\ }\textbf {\bibinfo {volume} {98}},\ \bibinfo
  {pages} {043535} (\bibinfo {year} {2018})},\ \Eprint
  {http://arxiv.org/abs/1805.10533} {arXiv:1805.10533 [hep-ph]} \BibitemShut
  {NoStop}%
\bibitem [{\citenamefont {Schwinger}(1966)}]{Schwinger:1966nj}%
  \BibitemOpen
  \bibfield  {author} {\bibinfo {author} {\bibfnamefont {J.~S.}\ \bibnamefont
  {Schwinger}},\ }\href {\doibase 10.1103/PhysRev.144.1087} {\bibfield
  {journal} {\bibinfo  {journal} {Phys. Rev.}\ }\textbf {\bibinfo {volume}
  {144}},\ \bibinfo {pages} {1087} (\bibinfo {year} {1966})}\BibitemShut
  {NoStop}%
\bibitem [{\citenamefont {Zwanziger}(1968)}]{Zwanziger:1968rs}%
  \BibitemOpen
  \bibfield  {author} {\bibinfo {author} {\bibfnamefont {D.}~\bibnamefont
  {Zwanziger}},\ }\href {\doibase 10.1103/PhysRev.176.1489} {\bibfield
  {journal} {\bibinfo  {journal} {Phys. Rev.}\ }\textbf {\bibinfo {volume}
  {176}},\ \bibinfo {pages} {1489} (\bibinfo {year} {1968})}\BibitemShut
  {NoStop}%
\bibitem [{\citenamefont {Zwanziger}(1971)}]{Zwanziger:1970hk}%
  \BibitemOpen
  \bibfield  {author} {\bibinfo {author} {\bibfnamefont {D.}~\bibnamefont
  {Zwanziger}},\ }\href {\doibase 10.1103/PhysRevD.3.880} {\bibfield  {journal}
  {\bibinfo  {journal} {Phys. Rev. D}\ }\textbf {\bibinfo {volume} {3}},\
  \bibinfo {pages} {880} (\bibinfo {year} {1971})}\BibitemShut {NoStop}%
\bibitem [{\citenamefont {Sokolov}\ and\ \citenamefont
  {Ringwald}(2022)}]{Sokolov:2022fvs}%
  \BibitemOpen
  \bibfield  {author} {\bibinfo {author} {\bibfnamefont {A.~V.}\ \bibnamefont
  {Sokolov}}\ and\ \bibinfo {author} {\bibfnamefont {A.}~\bibnamefont
  {Ringwald}},\ }\href@noop {} {\  (\bibinfo {year} {2022})},\ \Eprint
  {http://arxiv.org/abs/2205.02605} {arXiv:2205.02605 [hep-ph]} \BibitemShut
  {NoStop}%
\bibitem [{\citenamefont {Sikivie}(1983)}]{Sikivie:1983ip}%
  \BibitemOpen
  \bibfield  {author} {\bibinfo {author} {\bibfnamefont {P.}~\bibnamefont
  {Sikivie}},\ }\href {\doibase 10.1103/PhysRevLett.51.1415} {\bibfield
  {journal} {\bibinfo  {journal} {Phys. Rev. Lett.}\ }\textbf {\bibinfo
  {volume} {51}},\ \bibinfo {pages} {1415} (\bibinfo {year} {1983})},\ \bibinfo
  {note} {[Erratum: Phys.Rev.Lett. 52, 695 (1984)]}\BibitemShut {NoStop}%
\bibitem [{\citenamefont {Li}\ \emph {et~al.}(2023{\natexlab{a}})\citenamefont
  {Li}, \citenamefont {Zhang},\ and\ \citenamefont {Dai}}]{Li:2022oel}%
  \BibitemOpen
  \bibfield  {author} {\bibinfo {author} {\bibfnamefont {T.}~\bibnamefont
  {Li}}, \bibinfo {author} {\bibfnamefont {R.-J.}\ \bibnamefont {Zhang}}, \
  and\ \bibinfo {author} {\bibfnamefont {C.-J.}\ \bibnamefont {Dai}},\ }\href
  {\doibase 10.1007/JHEP03(2023)088} {\bibfield  {journal} {\bibinfo  {journal}
  {JHEP}\ }\textbf {\bibinfo {volume} {03}},\ \bibinfo {pages} {088} (\bibinfo
  {year} {2023}{\natexlab{a}})},\ \Eprint {http://arxiv.org/abs/2211.06847}
  {arXiv:2211.06847 [hep-ph]} \BibitemShut {NoStop}%
\bibitem [{\citenamefont {Tobar}\ \emph {et~al.}(2022)\citenamefont {Tobar},
  \citenamefont {Thomson}, \citenamefont {McAllister}, \citenamefont
  {Goryachev}, \citenamefont {Sokolov},\ and\ \citenamefont
  {Ringwald}}]{Tobar:2022rko}%
  \BibitemOpen
  \bibfield  {author} {\bibinfo {author} {\bibfnamefont {M.~E.}\ \bibnamefont
  {Tobar}}, \bibinfo {author} {\bibfnamefont {C.~A.}\ \bibnamefont {Thomson}},
  \bibinfo {author} {\bibfnamefont {B.~T.}\ \bibnamefont {McAllister}},
  \bibinfo {author} {\bibfnamefont {M.}~\bibnamefont {Goryachev}}, \bibinfo
  {author} {\bibfnamefont {A.}~\bibnamefont {Sokolov}}, \ and\ \bibinfo
  {author} {\bibfnamefont {A.}~\bibnamefont {Ringwald}},\ }\href@noop {} {\
  (\bibinfo {year} {2022})},\ \Eprint {http://arxiv.org/abs/2211.09637}
  {arXiv:2211.09637 [hep-ph]} \BibitemShut {NoStop}%
\bibitem [{\citenamefont {McAllister}\ \emph {et~al.}(2022)\citenamefont
  {McAllister}, \citenamefont {Quiskamp}, \citenamefont {O'Hare}, \citenamefont
  {Altin}, \citenamefont {Ivanov}, \citenamefont {Goryachev},\ and\
  \citenamefont {Tobar}}]{McAllister:2022ibe}%
  \BibitemOpen
  \bibfield  {author} {\bibinfo {author} {\bibfnamefont {B.~T.}\ \bibnamefont
  {McAllister}}, \bibinfo {author} {\bibfnamefont {A.}~\bibnamefont
  {Quiskamp}}, \bibinfo {author} {\bibfnamefont {C.}~\bibnamefont {O'Hare}},
  \bibinfo {author} {\bibfnamefont {P.}~\bibnamefont {Altin}}, \bibinfo
  {author} {\bibfnamefont {E.}~\bibnamefont {Ivanov}}, \bibinfo {author}
  {\bibfnamefont {M.}~\bibnamefont {Goryachev}}, \ and\ \bibinfo {author}
  {\bibfnamefont {M.}~\bibnamefont {Tobar}},\ }\href@noop {} {\  (\bibinfo
  {year} {2022})},\ \Eprint {http://arxiv.org/abs/2212.01971} {arXiv:2212.01971
  [hep-ph]} \BibitemShut {NoStop}%
\bibitem [{\citenamefont {Li}\ \emph {et~al.}(2023{\natexlab{b}})\citenamefont
  {Li}, \citenamefont {Dai},\ and\ \citenamefont {Zhang}}]{Li:2023kfh}%
  \BibitemOpen
  \bibfield  {author} {\bibinfo {author} {\bibfnamefont {T.}~\bibnamefont
  {Li}}, \bibinfo {author} {\bibfnamefont {C.-J.}\ \bibnamefont {Dai}}, \ and\
  \bibinfo {author} {\bibfnamefont {R.-J.}\ \bibnamefont {Zhang}},\ }\href@noop
  {} {\  (\bibinfo {year} {2023}{\natexlab{b}})},\ \Eprint
  {http://arxiv.org/abs/2304.12525} {arXiv:2304.12525 [hep-ph]} \BibitemShut
  {NoStop}%
\bibitem [{\citenamefont {Brandt}\ \emph {et~al.}(1978)\citenamefont {Brandt},
  \citenamefont {Neri},\ and\ \citenamefont {Zwanziger}}]{Brandt:1977be}%
  \BibitemOpen
  \bibfield  {author} {\bibinfo {author} {\bibfnamefont {R.~A.}\ \bibnamefont
  {Brandt}}, \bibinfo {author} {\bibfnamefont {F.}~\bibnamefont {Neri}}, \ and\
  \bibinfo {author} {\bibfnamefont {D.}~\bibnamefont {Zwanziger}},\ }\href
  {\doibase 10.1103/PhysRevLett.40.147} {\bibfield  {journal} {\bibinfo
  {journal} {Phys. Rev. Lett.}\ }\textbf {\bibinfo {volume} {40}},\ \bibinfo
  {pages} {147} (\bibinfo {year} {1978})}\BibitemShut {NoStop}%
\bibitem [{\citenamefont {Brandt}\ \emph {et~al.}(1979)\citenamefont {Brandt},
  \citenamefont {Neri},\ and\ \citenamefont {Zwanziger}}]{Brandt:1978wc}%
  \BibitemOpen
  \bibfield  {author} {\bibinfo {author} {\bibfnamefont {R.~A.}\ \bibnamefont
  {Brandt}}, \bibinfo {author} {\bibfnamefont {F.}~\bibnamefont {Neri}}, \ and\
  \bibinfo {author} {\bibfnamefont {D.}~\bibnamefont {Zwanziger}},\ }\href
  {\doibase 10.1103/PhysRevD.19.1153} {\bibfield  {journal} {\bibinfo
  {journal} {Phys. Rev. D}\ }\textbf {\bibinfo {volume} {19}},\ \bibinfo
  {pages} {1153} (\bibinfo {year} {1979})}\BibitemShut {NoStop}%
\bibitem [{\citenamefont {Sokolov}\ and\ \citenamefont
  {Ringwald}(2023)}]{Sokolov:2023pos}%
  \BibitemOpen
  \bibfield  {author} {\bibinfo {author} {\bibfnamefont {A.~V.}\ \bibnamefont
  {Sokolov}}\ and\ \bibinfo {author} {\bibfnamefont {A.}~\bibnamefont
  {Ringwald}},\ }\href@noop {} {\  (\bibinfo {year} {2023})},\ \Eprint
  {http://arxiv.org/abs/2303.10170} {arXiv:2303.10170 [hep-ph]} \BibitemShut
  {NoStop}%
\bibitem [{\citenamefont {Di~Luzio}\ \emph {et~al.}(2020)\citenamefont
  {Di~Luzio}, \citenamefont {Giannotti}, \citenamefont {Nardi},\ and\
  \citenamefont {Visinelli}}]{DiLuzio:2020wdo}%
  \BibitemOpen
  \bibfield  {author} {\bibinfo {author} {\bibfnamefont {L.}~\bibnamefont
  {Di~Luzio}}, \bibinfo {author} {\bibfnamefont {M.}~\bibnamefont {Giannotti}},
  \bibinfo {author} {\bibfnamefont {E.}~\bibnamefont {Nardi}}, \ and\ \bibinfo
  {author} {\bibfnamefont {L.}~\bibnamefont {Visinelli}},\ }\href {\doibase
  10.1016/j.physrep.2020.06.002} {\bibfield  {journal} {\bibinfo  {journal}
  {Phys. Rept.}\ }\textbf {\bibinfo {volume} {870}},\ \bibinfo {pages} {1}
  (\bibinfo {year} {2020})},\ \Eprint {http://arxiv.org/abs/2003.01100}
  {arXiv:2003.01100 [hep-ph]} \BibitemShut {NoStop}%
\bibitem [{\citenamefont {Preskill}\ \emph {et~al.}(1983)\citenamefont
  {Preskill}, \citenamefont {Wise},\ and\ \citenamefont
  {Wilczek}}]{Preskill:1982cy}%
  \BibitemOpen
  \bibfield  {author} {\bibinfo {author} {\bibfnamefont {J.}~\bibnamefont
  {Preskill}}, \bibinfo {author} {\bibfnamefont {M.~B.}\ \bibnamefont {Wise}},
  \ and\ \bibinfo {author} {\bibfnamefont {F.}~\bibnamefont {Wilczek}},\ }\href
  {\doibase 10.1016/0370-2693(83)90637-8} {\bibfield  {journal} {\bibinfo
  {journal} {Phys. Lett. B}\ }\textbf {\bibinfo {volume} {120}},\ \bibinfo
  {pages} {127} (\bibinfo {year} {1983})}\BibitemShut {NoStop}%
\bibitem [{\citenamefont {Dine}\ and\ \citenamefont
  {Fischler}(1983)}]{Dine:1982ah}%
  \BibitemOpen
  \bibfield  {author} {\bibinfo {author} {\bibfnamefont {M.}~\bibnamefont
  {Dine}}\ and\ \bibinfo {author} {\bibfnamefont {W.}~\bibnamefont
  {Fischler}},\ }\href {\doibase 10.1016/0370-2693(83)90639-1} {\bibfield
  {journal} {\bibinfo  {journal} {Phys. Lett. B}\ }\textbf {\bibinfo {volume}
  {120}},\ \bibinfo {pages} {137} (\bibinfo {year} {1983})}\BibitemShut
  {NoStop}%
\bibitem [{\citenamefont {Du}\ \emph {et~al.}(2018)\citenamefont {Du} \emph
  {et~al.}}]{ADMX:2018gho}%
  \BibitemOpen
  \bibfield  {author} {\bibinfo {author} {\bibfnamefont {N.}~\bibnamefont {Du}}
  \emph {et~al.} (\bibinfo {collaboration} {ADMX}),\ }\href {\doibase
  10.1103/PhysRevLett.120.151301} {\bibfield  {journal} {\bibinfo  {journal}
  {Phys. Rev. Lett.}\ }\textbf {\bibinfo {volume} {120}},\ \bibinfo {pages}
  {151301} (\bibinfo {year} {2018})},\ \Eprint
  {http://arxiv.org/abs/1804.05750} {arXiv:1804.05750 [hep-ex]} \BibitemShut
  {NoStop}%
\bibitem [{\citenamefont {Bartram}\ \emph {et~al.}(2021)\citenamefont {Bartram}
  \emph {et~al.}}]{ADMX:2021nhd}%
  \BibitemOpen
  \bibfield  {author} {\bibinfo {author} {\bibfnamefont {C.}~\bibnamefont
  {Bartram}} \emph {et~al.} (\bibinfo {collaboration} {ADMX}),\ }\href
  {\doibase 10.1103/PhysRevLett.127.261803} {\bibfield  {journal} {\bibinfo
  {journal} {Phys. Rev. Lett.}\ }\textbf {\bibinfo {volume} {127}},\ \bibinfo
  {pages} {261803} (\bibinfo {year} {2021})},\ \Eprint
  {http://arxiv.org/abs/2110.06096} {arXiv:2110.06096 [hep-ex]} \BibitemShut
  {NoStop}%
\bibitem [{\citenamefont {Sikivie}(2021)}]{Sikivie:2020zpn}%
  \BibitemOpen
  \bibfield  {author} {\bibinfo {author} {\bibfnamefont {P.}~\bibnamefont
  {Sikivie}},\ }\href {\doibase 10.1103/RevModPhys.93.015004} {\bibfield
  {journal} {\bibinfo  {journal} {Rev. Mod. Phys.}\ }\textbf {\bibinfo {volume}
  {93}},\ \bibinfo {pages} {015004} (\bibinfo {year} {2021})},\ \Eprint
  {http://arxiv.org/abs/2003.02206} {arXiv:2003.02206 [hep-ph]} \BibitemShut
  {NoStop}%
\bibitem [{\citenamefont {Beutter}\ \emph {et~al.}(2019)\citenamefont
  {Beutter}, \citenamefont {Pargner}, \citenamefont {Schwetz},\ and\
  \citenamefont {Todarello}}]{Beutter:2018xfx}%
  \BibitemOpen
  \bibfield  {author} {\bibinfo {author} {\bibfnamefont {M.}~\bibnamefont
  {Beutter}}, \bibinfo {author} {\bibfnamefont {A.}~\bibnamefont {Pargner}},
  \bibinfo {author} {\bibfnamefont {T.}~\bibnamefont {Schwetz}}, \ and\
  \bibinfo {author} {\bibfnamefont {E.}~\bibnamefont {Todarello}},\ }\href
  {\doibase 10.1088/1475-7516/2019/02/026} {\bibfield  {journal} {\bibinfo
  {journal} {JCAP}\ }\textbf {\bibinfo {volume} {02}},\ \bibinfo {pages} {026}
  (\bibinfo {year} {2019})},\ \Eprint {http://arxiv.org/abs/1812.05487}
  {arXiv:1812.05487 [hep-ph]} \BibitemShut {NoStop}%
\bibitem [{\citenamefont {Yang}\ \emph {et~al.}(2022)\citenamefont {Yang},
  \citenamefont {Gao},\ and\ \citenamefont {Peng}}]{Yang:2022uil}%
  \BibitemOpen
  \bibfield  {author} {\bibinfo {author} {\bibfnamefont {Q.}~\bibnamefont
  {Yang}}, \bibinfo {author} {\bibfnamefont {Y.}~\bibnamefont {Gao}}, \ and\
  \bibinfo {author} {\bibfnamefont {Z.}~\bibnamefont {Peng}},\ }\href@noop {}
  {\  (\bibinfo {year} {2022})},\ \Eprint {http://arxiv.org/abs/2201.08291}
  {arXiv:2201.08291 [hep-ph]} \BibitemShut {NoStop}%
\bibitem [{\citenamefont {Donohue}\ \emph {et~al.}(2021)\citenamefont
  {Donohue}, \citenamefont {Gardner},\ and\ \citenamefont
  {Korsch}}]{Donohue:2021jbv}%
  \BibitemOpen
  \bibfield  {author} {\bibinfo {author} {\bibfnamefont {C.~M.}\ \bibnamefont
  {Donohue}}, \bibinfo {author} {\bibfnamefont {S.}~\bibnamefont {Gardner}}, \
  and\ \bibinfo {author} {\bibfnamefont {W.}~\bibnamefont {Korsch}},\
  }\href@noop {} {\  (\bibinfo {year} {2021})},\ \Eprint
  {http://arxiv.org/abs/2109.08163} {arXiv:2109.08163 [hep-ph]} \BibitemShut
  {NoStop}%
\bibitem [{\citenamefont {Fujikawa}(1979)}]{Fujikawa:1979ay}%
  \BibitemOpen
  \bibfield  {author} {\bibinfo {author} {\bibfnamefont {K.}~\bibnamefont
  {Fujikawa}},\ }\href {\doibase 10.1103/PhysRevLett.42.1195} {\bibfield
  {journal} {\bibinfo  {journal} {Phys. Rev. Lett.}\ }\textbf {\bibinfo
  {volume} {42}},\ \bibinfo {pages} {1195} (\bibinfo {year}
  {1979})}\BibitemShut {NoStop}%
\bibitem [{\citenamefont {Kapustin}\ and\ \citenamefont
  {Seiberg}(2014)}]{Kapustin:2014gua}%
  \BibitemOpen
  \bibfield  {author} {\bibinfo {author} {\bibfnamefont {A.}~\bibnamefont
  {Kapustin}}\ and\ \bibinfo {author} {\bibfnamefont {N.}~\bibnamefont
  {Seiberg}},\ }\href {\doibase 10.1007/JHEP04(2014)001} {\bibfield  {journal}
  {\bibinfo  {journal} {JHEP}\ }\textbf {\bibinfo {volume} {04}},\ \bibinfo
  {pages} {001} (\bibinfo {year} {2014})},\ \Eprint
  {http://arxiv.org/abs/1401.0740} {arXiv:1401.0740 [hep-th]} \BibitemShut
  {NoStop}%
\bibitem [{\citenamefont {Seiberg}(2010)}]{Seiberg:2010qd}%
  \BibitemOpen
  \bibfield  {author} {\bibinfo {author} {\bibfnamefont {N.}~\bibnamefont
  {Seiberg}},\ }\href {\doibase 10.1007/JHEP07(2010)070} {\bibfield  {journal}
  {\bibinfo  {journal} {JHEP}\ }\textbf {\bibinfo {volume} {07}},\ \bibinfo
  {pages} {070} (\bibinfo {year} {2010})},\ \Eprint
  {http://arxiv.org/abs/1005.0002} {arXiv:1005.0002 [hep-th]} \BibitemShut
  {NoStop}%
\bibitem [{\citenamefont {Gaiotto}\ \emph {et~al.}(2015)\citenamefont
  {Gaiotto}, \citenamefont {Kapustin}, \citenamefont {Seiberg},\ and\
  \citenamefont {Willett}}]{Gaiotto:2014kfa}%
  \BibitemOpen
  \bibfield  {author} {\bibinfo {author} {\bibfnamefont {D.}~\bibnamefont
  {Gaiotto}}, \bibinfo {author} {\bibfnamefont {A.}~\bibnamefont {Kapustin}},
  \bibinfo {author} {\bibfnamefont {N.}~\bibnamefont {Seiberg}}, \ and\
  \bibinfo {author} {\bibfnamefont {B.}~\bibnamefont {Willett}},\ }\href
  {\doibase 10.1007/JHEP02(2015)172} {\bibfield  {journal} {\bibinfo  {journal}
  {JHEP}\ }\textbf {\bibinfo {volume} {02}},\ \bibinfo {pages} {172} (\bibinfo
  {year} {2015})},\ \Eprint {http://arxiv.org/abs/1412.5148} {arXiv:1412.5148
  [hep-th]} \BibitemShut {NoStop}%
\bibitem [{\citenamefont {Brennan}\ \emph {et~al.}(2023)\citenamefont
  {Brennan}, \citenamefont {Hong},\ and\ \citenamefont
  {Wang}}]{Brennan:2023kpw}%
  \BibitemOpen
  \bibfield  {author} {\bibinfo {author} {\bibfnamefont {T.~D.}\ \bibnamefont
  {Brennan}}, \bibinfo {author} {\bibfnamefont {S.}~\bibnamefont {Hong}}, \
  and\ \bibinfo {author} {\bibfnamefont {L.-T.}\ \bibnamefont {Wang}},\
  }\href@noop {} {\  (\bibinfo {year} {2023})},\ \Eprint
  {http://arxiv.org/abs/2302.00777} {arXiv:2302.00777 [hep-ph]} \BibitemShut
  {NoStop}%
\bibitem [{\citenamefont {Krauss}\ \emph {et~al.}(1985)\citenamefont {Krauss},
  \citenamefont {Moody}, \citenamefont {Wilczek},\ and\ \citenamefont
  {Morris}}]{Krauss:1985ub}%
  \BibitemOpen
  \bibfield  {author} {\bibinfo {author} {\bibfnamefont {L.}~\bibnamefont
  {Krauss}}, \bibinfo {author} {\bibfnamefont {J.}~\bibnamefont {Moody}},
  \bibinfo {author} {\bibfnamefont {F.}~\bibnamefont {Wilczek}}, \ and\
  \bibinfo {author} {\bibfnamefont {D.~E.}\ \bibnamefont {Morris}},\ }\href
  {\doibase 10.1103/PhysRevLett.55.1797} {\bibfield  {journal} {\bibinfo
  {journal} {Phys. Rev. Lett.}\ }\textbf {\bibinfo {volume} {55}},\ \bibinfo
  {pages} {1797} (\bibinfo {year} {1985})}\BibitemShut {NoStop}%
\bibitem [{\citenamefont {Ioannisian}\ \emph {et~al.}(2017)\citenamefont
  {Ioannisian}, \citenamefont {Kazarian}, \citenamefont {Millar},\ and\
  \citenamefont {Raffelt}}]{Ioannisian:2017srr}%
  \BibitemOpen
  \bibfield  {author} {\bibinfo {author} {\bibfnamefont {A.~N.}\ \bibnamefont
  {Ioannisian}}, \bibinfo {author} {\bibfnamefont {N.}~\bibnamefont
  {Kazarian}}, \bibinfo {author} {\bibfnamefont {A.~J.}\ \bibnamefont
  {Millar}}, \ and\ \bibinfo {author} {\bibfnamefont {G.~G.}\ \bibnamefont
  {Raffelt}},\ }\href {\doibase 10.1088/1475-7516/2017/09/005} {\bibfield
  {journal} {\bibinfo  {journal} {JCAP}\ }\textbf {\bibinfo {volume} {09}},\
  \bibinfo {pages} {005} (\bibinfo {year} {2017})},\ \Eprint
  {http://arxiv.org/abs/1707.00701} {arXiv:1707.00701 [hep-ph]} \BibitemShut
  {NoStop}%
\bibitem [{\citenamefont {Workman}\ and\ \citenamefont
  {Others}(2022)}]{Workman:2022ynf}%
  \BibitemOpen
  \bibfield  {author} {\bibinfo {author} {\bibfnamefont {R.~L.}\ \bibnamefont
  {Workman}}\ and\ \bibinfo {author} {\bibnamefont {Others}} (\bibinfo
  {collaboration} {Particle Data Group}),\ }\href {\doibase
  10.1093/ptep/ptac097} {\bibfield  {journal} {\bibinfo  {journal} {PTEP}\
  }\textbf {\bibinfo {volume} {2022}},\ \bibinfo {pages} {083C01} (\bibinfo
  {year} {2022})}\BibitemShut {NoStop}%
\bibitem [{\citenamefont {Hagmann}\ \emph {et~al.}(1990)\citenamefont
  {Hagmann}, \citenamefont {Sikivie}, \citenamefont {Sullivan},\ and\
  \citenamefont {Tanner}}]{Hagmann:1990tj}%
  \BibitemOpen
  \bibfield  {author} {\bibinfo {author} {\bibfnamefont {C.}~\bibnamefont
  {Hagmann}}, \bibinfo {author} {\bibfnamefont {P.}~\bibnamefont {Sikivie}},
  \bibinfo {author} {\bibfnamefont {N.~S.}\ \bibnamefont {Sullivan}}, \ and\
  \bibinfo {author} {\bibfnamefont {D.~B.}\ \bibnamefont {Tanner}},\ }\href
  {\doibase 10.1103/PhysRevD.42.1297} {\bibfield  {journal} {\bibinfo
  {journal} {Phys. Rev. D}\ }\textbf {\bibinfo {volume} {42}},\ \bibinfo
  {pages} {1297} (\bibinfo {year} {1990})}\BibitemShut {NoStop}%
\bibitem [{\citenamefont {Anastassopoulos}\ \emph {et~al.}(2017)\citenamefont
  {Anastassopoulos} \emph {et~al.}}]{CAST:2017uph}%
  \BibitemOpen
  \bibfield  {author} {\bibinfo {author} {\bibfnamefont {V.}~\bibnamefont
  {Anastassopoulos}} \emph {et~al.} (\bibinfo {collaboration} {CAST}),\ }\href
  {\doibase 10.1038/nphys4109} {\bibfield  {journal} {\bibinfo  {journal}
  {Nature Phys.}\ }\textbf {\bibinfo {volume} {13}},\ \bibinfo {pages} {584}
  (\bibinfo {year} {2017})},\ \Eprint {http://arxiv.org/abs/1705.02290}
  {arXiv:1705.02290 [hep-ex]} \BibitemShut {NoStop}%
\bibitem [{\citenamefont {Crisosto}\ \emph {et~al.}(2020)\citenamefont
  {Crisosto}, \citenamefont {Sikivie}, \citenamefont {Sullivan}, \citenamefont
  {Tanner}, \citenamefont {Yang},\ and\ \citenamefont
  {Rybka}}]{Crisosto:2019fcj}%
  \BibitemOpen
  \bibfield  {author} {\bibinfo {author} {\bibfnamefont {N.}~\bibnamefont
  {Crisosto}}, \bibinfo {author} {\bibfnamefont {P.}~\bibnamefont {Sikivie}},
  \bibinfo {author} {\bibfnamefont {N.~S.}\ \bibnamefont {Sullivan}}, \bibinfo
  {author} {\bibfnamefont {D.~B.}\ \bibnamefont {Tanner}}, \bibinfo {author}
  {\bibfnamefont {J.}~\bibnamefont {Yang}}, \ and\ \bibinfo {author}
  {\bibfnamefont {G.}~\bibnamefont {Rybka}},\ }\href {\doibase
  10.1103/PhysRevLett.124.241101} {\bibfield  {journal} {\bibinfo  {journal}
  {Phys. Rev. Lett.}\ }\textbf {\bibinfo {volume} {124}},\ \bibinfo {pages}
  {241101} (\bibinfo {year} {2020})},\ \Eprint
  {http://arxiv.org/abs/1911.05772} {arXiv:1911.05772 [astro-ph.CO]}
  \BibitemShut {NoStop}%
\end{thebibliography}%

\end{document}